\documentclass[draft,onecolumn]{IEEEtran}

\usepackage[utf8]{inputenc} 
\usepackage[T1]{fontenc}
\usepackage{url}
\usepackage{xcolor}
\usepackage{ifthen}
\usepackage{cite}
\usepackage{comment}
\usepackage[cmex10]{amsmath} % Use the [cmex10] option to ensure complicance
\usepackage{amssymb}
\usepackage{amsthm}    
\usepackage{graphicx}
\usepackage{subcaption}

\usepackage{etoolbox}
\makeatletter
\patchcmd{\@makecaption}
  {\scshape}
  {}
  {}
  {}
\makeatletter
\patchcmd{\@makecaption}
  {\\}
  {.\ }
  {}
  {}
\makeatother
\def\tablename{Table}

\usepackage{amsmath}
\usepackage{tikz}
\usepackage{mathdots}
\usepackage{yhmath}
\usepackage{cancel}
\usepackage{color}
\usepackage{siunitx}
\usepackage{array}
\usepackage{multirow}
\usepackage{hyperref}
\usepackage{amssymb}
\usepackage{gensymb}
\usepackage{tabularx}
\usepackage{booktabs}
\usetikzlibrary{fadings}
\usetikzlibrary{patterns}
\usetikzlibrary{shadows.blur}

\usepackage{pgf,pgfplots}
\pgfplotsset{compat=1.14}
\usepackage{mathrsfs}
\usetikzlibrary{arrows}
\definecolor{rvwvcq}{rgb}{0.08235294117647059,0.396078431372549,0.7529411764705882}

\theoremstyle{definition}
\newtheorem*{definition}{Definition}

\newtheorem{theorem}{Theorem}

\newtheorem{lemma}{Lemma}
\newtheorem{proposition}{Proposition}

\usepackage[ruled,linesnumbered,vlined]{algorithm2e}
\renewcommand{\footnotesize}{\small}

\begin{document}
\title{Fundamental Limits of Distributed Encoding} 

% %%% Single author, or several authors with same affiliation:
% \author{%
%   \IEEEauthorblockN{Stefan M.~Moser}
%   \IEEEauthorblockA{ETH Zürich\\
%                     ISI (D-ITET)\\
%                     CH-8092 Zürich, Switzerland\\
%                     Email: moser@isi.ee.ethz.ch}
% }

%%% Several authors with up to three affiliations:
\author{Nastaran Abadi Khooshemehr, Mohammad Ali Maddah-Ali \\
Department of Electrical Engineering, Sharif University of Technology\\
abadi.nastaran@ee.sharif.edu, maddah\_ali@sharif.edu }

% \renewcommand{\footnotesize}{\footnotesize }

%%% Many authors with many affiliations:
% \author{%
%   \IEEEauthorblockN{Albus Dumbledore\IEEEauthorrefmark{1},
%                     Olympe Maxime\IEEEauthorrefmark{2},
%                     Stefan M.~Moser\IEEEauthorrefmark{3}\IEEEauthorrefmark{4},
%                     and Harry Potter\IEEEauthorrefmark{1}}
%   \IEEEauthorblockA{\IEEEauthorrefmark{1}%
%                     Hogwarts School of Witchcraft and Wizardry,
%                     1714 Hogsmeade, Scotland,
%                     \{dumbledore, potter\}@hogwarts.edu}
%   \IEEEauthorblockA{\IEEEauthorrefmark{2}%
%                     Beauxbatons Academy of Magic,
%                     1290 Pyrénées, France,
%                     maxime@beauxbatons.edu}
%   \IEEEauthorblockA{\IEEEauthorrefmark{3}%
%                     ETH Zürich, ISI (D-ITET), ETH Zentrum, 
%                     CH-8092 Zürich, Switzerland,
%                     moser@isi.ee.ethz.ch}
%   \IEEEauthorblockA{\IEEEauthorrefmark{4}%
%                     National Chiao Tung University (NCTU), 
%                     Hsinchu, Taiwan,
%                     moser@isi.ee.ethz.ch}
% }

\maketitle

\begin{abstract}
In general coding theory, we often assume that error is observed in transferring or storing encoded symbols, while the process of encoding itself is error-free. Motivated by recent applications of coding theory, in this paper, we consider the case where the process of encoding is distributed and prone to error. We introduce the problem of distributed encoding,  comprising of a set of $K \in \mathbb{N}$  isolated source nodes and $N \in \mathbb{N}$ encoding nodes.  Each source node has one symbol from a finite field, which is sent to each of the encoding nodes. Each encoding node stores an encoded symbol from the same field, as a function of the received symbols. However, some of the source nodes are controlled by the adversary and may send different symbols to different encoding nodes.  Depending on the number of adversarial nodes, denoted by $\beta \in \mathbb{N}$, and the cardinality of the set of symbols that each one generates, denoted by $v \in \mathbb{N}$, the process of decoding from the encoded symbols could be impossible. Assume that a decoder connects to an arbitrary subset of $t \in \mathbb{N}$ encoding nodes and wants to decode the symbols of the honest nodes correctly, without necessarily identifying the sets of honest and adversarial nodes. In this paper, we study $t^* \in \mathbb{N}$, the minimum of such $t$, which is a function of $K$, $N$, $\beta$, and $v$. We show that
   when the encoding nodes use linear coding, $t^*_{\textrm{linear}}=K+2\beta(v-1)$, if $N\ge K+2\beta(v-1)$, and $t^*_{\textrm{linear}}=N$, if $N\le K+2\beta(v-1)$.
   In order to achieve $t^*_{\textrm{linear}}$, we use random linear coding and show that in any feasible solution that the decoder finds, the messages of the honest nodes are decoded correctly.
   In order to prove the converse of the fundamental limit, we show that when the adversary behaves in a particular way, it can always confuse the decoder between two feasible that differ in the message of at least one honest node.
\end{abstract}

%% The paper must be self-contained. However, if you are referring to
%% a full version for checking certain proofs, please provide the
%% publically accessible location below.  If the paper is completely
%% self-contained, you can remove the following line from your
%% submission.

\section{Introduction}
In general coding theory, it is often assumed that the encoder is immune to errors of any kind. In other words, errors affect the coded symbols, after the intact process of encoding. Recently, new scenarios have emerged in which the sources of data are distributed. Actually, these days we deal with huge amounts of data whose production is distributed, as in IoT and blockchain applications. To protect such data, we store encoded versions of them in a distributed fashion. Thus, the source and the encoding of data are both distributed. An example of such scenario is \emph{sharding} in blockchain \cite{li2018polyshard,shardingFAQ}. In distributed encoding of distributed data, the basic assumption of error-free encoding might no longer be valid.

The use of coding in distributed systems is motivated by its successful application in communication systems. For example, \cite{dimakis2011survey} reviews coding for distributed storage systems. In communication systems, coding is used to protect data against errors or erasures caused by the communication channel. There are two approaches to model the error in channels and find the fundamental limits of coding. \emph{Shannon} considered a probabilistic model for errors and defined successful communication as one in which the probability of decoding error approaches zero, as the length of the code increases. With that model, the notion of the \emph{capacity} of the channel determines the fundamental limits on the rate of coding in successful communication.  On the other hand, \emph{Hamming} took the \emph{worst-case} approach, in which up to a certain number of errors might happen in any locations, chosen by an \emph{adversary}. Therefore an appropriate code should be able to correct errors of any pattern.
For those scenarios, the fundamental limits of coding characterize a region for the parameters of the code, e.g. minimum distance, size of the code, length of the message, and codeword vectors, where one can guarantee that an error correcting code with those parameters exists. On the other side, if the parameters are out of that region, it is certain that no code with those parameters exists. Characterizing fundamental limits in the second model usually leads to combinatorial problems. Examples of fundamental inequalities in coding theory are the well-known Singleton bound, Plotkin bound and Johnson bound, See \cite{ling2004coding}.

\begin{comment}
The specific case of our interest is the presence of adversarial source nodes that give inconsistent data to the different encoding entities. This can make the process of decoding impossible. 
For instance, \cite{dimakis2010network,rashmi2012regenerating,pawar2011securing,pawar2011securing2,kosut2013polytope,bitar2015securing} make such fundamental assumption.
\end{comment}

In this work, we investigate the problem of distributed encoding and storing in the presence of adversaries. We assume there are $K \in \mathbb{N}$ isolated source nodes that produce data \textit{symbols}. The source nodes are connected to $N\in \mathbb{N}$ nodes, called encoding nodes. Each encoding node stores a (coded) symbol as a  function of the symbols it receives, so that the original symbols are recoverable from the coded symbols. The challenge is that some of the source nodes, up to $\beta \in \mathbb{N}$, might be \textit{adversarial} and send different symbols to different nodes as an attack to prevent the correct recovery of the other symbols. The goal is to be able to correctly recover the input symbols of the \emph{honest} source nodes, using the coded values of the encoding nodes, in every possible scenario of adversarial acts. By honest source nodes, we mean those that act consistently and send the same data to all of the encoding nodes. By adversarial act, we mean sending different data to different encoding nodes. Note that we do not know the adversarial source nodes, and they can be any of the source nodes. We assume that the number of different symbols that a source node can inject into the system is upper limited by an integer $v$. This is a fair assumption because there are already methods, such as proof-of-work, in distributed systems that do not allow the source nodes to flood the system.
A decoder connects to an arbitrary subset of  $t \in \mathbb{N}$ encoding nodes and decodes the messages. In this paper, we characterize $t^*$ as the minimum of $t$, for which there are encoding functions such that the decoder can recover the messages of the honest nodes correctly.

\begin{figure}
    \centering
\begin{tikzpicture}[x=0.75pt,y=0.75pt,yscale=-1,xscale=1, scale=0.9]
%uncomment if require: \path (0,647); %set diagram left start at 0, and has height of 647

%Shape: Circle [id:dp7608668007570747] 
\draw  [fill={rgb, 255:red, 80; green, 227; blue, 194 }  ,fill opacity=1 ] (121,173) .. controls (121,159.19) and (132.19,148) .. (146,148) .. controls (159.81,148) and (171,159.19) .. (171,173) .. controls (171,186.81) and (159.81,198) .. (146,198) .. controls (132.19,198) and (121,186.81) .. (121,173) -- cycle ;
%Straight Lines [id:da6681639078501869] 
\draw    (170,172) -- (268.45,91.27) ;
\draw [shift={(270,90)}, rotate = 500.65] [color={rgb, 255:red, 0; green, 0; blue, 0 }  ][line width=0.75]    (10.93,-3.29) .. controls (6.95,-1.4) and (3.31,-0.3) .. (0,0) .. controls (3.31,0.3) and (6.95,1.4) .. (10.93,3.29)   ;
%Shape: Rectangle [id:dp0426732417549327] 
\draw  [fill={rgb, 255:red, 126; green, 211; blue, 33 }  ,fill opacity=1 ] (271,79.19) -- (400,79.19) -- (400,119.19) -- (271,119.19) -- cycle ;
%Straight Lines [id:da956661554724372] 
\draw    (170,172) -- (268.01,160.24) ;
\draw [shift={(270,160)}, rotate = 533.1600000000001] [color={rgb, 255:red, 0; green, 0; blue, 0 }  ][line width=0.75]    (10.93,-3.29) .. controls (6.95,-1.4) and (3.31,-0.3) .. (0,0) .. controls (3.31,0.3) and (6.95,1.4) .. (10.93,3.29)   ;
%Straight Lines [id:da20755573077182143] 
\draw    (170,172) -- (268.27,229) ;
\draw [shift={(270,230)}, rotate = 210.11] [color={rgb, 255:red, 0; green, 0; blue, 0 }  ][line width=0.75]    (10.93,-3.29) .. controls (6.95,-1.4) and (3.31,-0.3) .. (0,0) .. controls (3.31,0.3) and (6.95,1.4) .. (10.93,3.29)   ;
%Shape: Circle [id:dp9968939744863246] 
\draw  [fill={rgb, 255:red, 80; green, 227; blue, 194 }  ,fill opacity=1 ] (121,242) .. controls (121,228.19) and (132.19,217) .. (146,217) .. controls (159.81,217) and (171,228.19) .. (171,242) .. controls (171,255.81) and (159.81,267) .. (146,267) .. controls (132.19,267) and (121,255.81) .. (121,242) -- cycle ;
%Straight Lines [id:da8101396121106783] 
\draw    (171,242) -- (268.86,101.64) ;
\draw [shift={(270,100)}, rotate = 484.88] [color={rgb, 255:red, 0; green, 0; blue, 0 }  ][line width=0.75]    (10.93,-3.29) .. controls (6.95,-1.4) and (3.31,-0.3) .. (0,0) .. controls (3.31,0.3) and (6.95,1.4) .. (10.93,3.29)   ;
%Straight Lines [id:da487276786386339] 
\draw    (171,242) -- (268.38,171.18) ;
\draw [shift={(270,170)}, rotate = 503.97] [color={rgb, 255:red, 0; green, 0; blue, 0 }  ][line width=0.75]    (10.93,-3.29) .. controls (6.95,-1.4) and (3.31,-0.3) .. (0,0) .. controls (3.31,0.3) and (6.95,1.4) .. (10.93,3.29)   ;
%Straight Lines [id:da3908328045923928] 
\draw    (171,242) -- (268,240.04) ;
\draw [shift={(270,240)}, rotate = 538.8399999999999] [color={rgb, 255:red, 0; green, 0; blue, 0 }  ][line width=0.75]    (10.93,-3.29) .. controls (6.95,-1.4) and (3.31,-0.3) .. (0,0) .. controls (3.31,0.3) and (6.95,1.4) .. (10.93,3.29)   ;
%Shape: Rectangle [id:dp8083682724134169] 
\draw  [fill={rgb, 255:red, 126; green, 211; blue, 33 }  ,fill opacity=1 ] (271,149.19) -- (400,149.19) -- (400,189.19) -- (271,189.19) -- cycle ;
%Shape: Rectangle [id:dp3046728683250488] 
\draw  [fill={rgb, 255:red, 126; green, 211; blue, 33 }  ,fill opacity=1 ] (271,219.19) -- (400,219.19) -- (400,259.19) -- (271,259.19) -- cycle ;
%Shape: Rectangle [id:dp12968764260424193] 
\draw  [fill={rgb, 255:red, 126; green, 211; blue, 33 }  ,fill opacity=1 ] (271,291.19) -- (400,291.19) -- (400,331.19) -- (271,331.19) -- cycle ;
%Shape: Circle [id:dp852043179698831] 
\draw  [fill={rgb, 255:red, 80; green, 227; blue, 194 }  ,fill opacity=1 ] (121,312) .. controls (121,298.19) and (132.19,287) .. (146,287) .. controls (159.81,287) and (171,298.19) .. (171,312) .. controls (171,325.81) and (159.81,337) .. (146,337) .. controls (132.19,337) and (121,325.81) .. (121,312) -- cycle ;
%Straight Lines [id:da6641522158828881] 
\draw    (171,173) -- (268.77,298.42) ;
\draw [shift={(270,300)}, rotate = 232.06] [color={rgb, 255:red, 0; green, 0; blue, 0 }  ][line width=0.75]    (10.93,-3.29) .. controls (6.95,-1.4) and (3.31,-0.3) .. (0,0) .. controls (3.31,0.3) and (6.95,1.4) .. (10.93,3.29)   ;
%Straight Lines [id:da6845393175073675] 
\draw    (171,242) -- (268.35,308.87) ;
\draw [shift={(270,310)}, rotate = 214.48] [color={rgb, 255:red, 0; green, 0; blue, 0 }  ][line width=0.75]    (10.93,-3.29) .. controls (6.95,-1.4) and (3.31,-0.3) .. (0,0) .. controls (3.31,0.3) and (6.95,1.4) .. (10.93,3.29)   ;
%Straight Lines [id:da8788632362855169] 
\draw    (171,312) -- (269.12,111.8) ;
\draw [shift={(270,110)}, rotate = 476.11] [color={rgb, 255:red, 0; green, 0; blue, 0 }  ][line width=0.75]    (10.93,-3.29) .. controls (6.95,-1.4) and (3.31,-0.3) .. (0,0) .. controls (3.31,0.3) and (6.95,1.4) .. (10.93,3.29)   ;
%Straight Lines [id:da8715974484862681] 
\draw    (171,312) -- (268.8,181.6) ;
\draw [shift={(270,180)}, rotate = 486.87] [color={rgb, 255:red, 0; green, 0; blue, 0 }  ][line width=0.75]    (10.93,-3.29) .. controls (6.95,-1.4) and (3.31,-0.3) .. (0,0) .. controls (3.31,0.3) and (6.95,1.4) .. (10.93,3.29)   ;
%Straight Lines [id:da9468503496528085] 
\draw    (171,312) -- (268.3,251.06) ;
\draw [shift={(270,250)}, rotate = 507.94] [color={rgb, 255:red, 0; green, 0; blue, 0 }  ][line width=0.75]    (10.93,-3.29) .. controls (6.95,-1.4) and (3.31,-0.3) .. (0,0) .. controls (3.31,0.3) and (6.95,1.4) .. (10.93,3.29)   ;
%Straight Lines [id:da02165333155700999] 
\draw    (171,312) -- (268.01,319.84) ;
\draw [shift={(270,320)}, rotate = 184.62] [color={rgb, 255:red, 0; green, 0; blue, 0 }  ][line width=0.75]    (10.93,-3.29) .. controls (6.95,-1.4) and (3.31,-0.3) .. (0,0) .. controls (3.31,0.3) and (6.95,1.4) .. (10.93,3.29)   ;
%Shape: Rectangle [id:dp7205840329442379] 
\draw  [fill={rgb, 255:red, 126; green, 211; blue, 33 }  ,fill opacity=1 ] (271,361.19) -- (400,361.19) -- (400,401.19) -- (271,401.19) -- cycle ;
%Straight Lines [id:da977547272438531] 
\draw    (171,312) -- (269.51,399.86) ;
\draw [shift={(271,401.19)}, rotate = 221.73] [color={rgb, 255:red, 0; green, 0; blue, 0 }  ][line width=0.75]    (10.93,-3.29) .. controls (6.95,-1.4) and (3.31,-0.3) .. (0,0) .. controls (3.31,0.3) and (6.95,1.4) .. (10.93,3.29)   ;
%Straight Lines [id:da16979856457654874] 
\draw    (171,242) -- (268.89,388.34) ;
\draw [shift={(270,390)}, rotate = 236.22] [color={rgb, 255:red, 0; green, 0; blue, 0 }  ][line width=0.75]    (10.93,-3.29) .. controls (6.95,-1.4) and (3.31,-0.3) .. (0,0) .. controls (3.31,0.3) and (6.95,1.4) .. (10.93,3.29)   ;
%Straight Lines [id:da563163283898038] 
\draw    (171,173) -- (269.1,368.21) ;
\draw [shift={(270,370)}, rotate = 243.32] [color={rgb, 255:red, 0; green, 0; blue, 0 }  ][line width=0.75]    (10.93,-3.29) .. controls (6.95,-1.4) and (3.31,-0.3) .. (0,0) .. controls (3.31,0.3) and (6.95,1.4) .. (10.93,3.29)   ;

% Text Node
\draw (184.5,148) node   [align=left] {$\displaystyle x'_{1}$};
% Text Node
\draw (335.5,99.19) node   [align=left] {$\displaystyle y_{1} =f_{1}( x'_{1} ,x_{2} ,x_{3})$};
% Text Node
\draw (335.5,169.19) node   [align=left] {$\displaystyle y_{2} =f_{2}( x'_{1} ,x_{2} ,x_{3})$};
% Text Node
\draw (335.5,239.19) node   [align=left] {$\displaystyle y_{3} =f_{3}( x_{1} ,x_{2} ,x_{3})$};
% Text Node
\draw (196.5,163) node   [align=left] {$\displaystyle x'_{1}$};
% Text Node
\draw (176.5,218) node   [align=left] {$\displaystyle x_{2}$};
% Text Node
\draw (196.5,230) node   [align=left] {$\displaystyle x_{2}$};
% Text Node
\draw (188.5,258) node   [align=left] {$\displaystyle x_{2}$};
% Text Node
\draw (174,191.81) node   [align=left] {$\displaystyle x_{1}$};
% Text Node
\draw (146,172) node   [align=left] {$\displaystyle 1$};
% Text Node
\draw (146,242) node   [align=left] {$\displaystyle 2$};
% Text Node
\draw (335.5,311.19) node   [align=left] {$\displaystyle y_{4} =f_{4}( x_{1} ,x_{2} ,x_{3})$};
% Text Node
\draw (197.5,177) node   [align=left] {$\displaystyle x_{1}$};
% Text Node
\draw (173.5,263) node   [align=left] {$\displaystyle x_{2}$};
% Text Node
\draw (146,312) node   [align=left] {$\displaystyle 3$};
% Text Node
\draw (176.5,284) node   [align=left] {$\displaystyle x_{3}$};
% Text Node
\draw (193.5,289) node   [align=left] {$\displaystyle x_{3}$};
% Text Node
\draw (195.5,305) node   [align=left] {$\displaystyle x_{3}$};
% Text Node
\draw (182.5,330) node   [align=left] {$\displaystyle x_{3}$};
% Text Node
\draw (335.5,380.5) node   [align=left] {$\displaystyle y_{5} =f_{5}( x_{1} ,x_{2} ,x_{3})$};
% Text Node
\draw (193.5,190) node   [align=left] {$\displaystyle x_{1}$};
% Text Node
\draw (190.5,245) node   [align=left] {$\displaystyle x_{2}$};
% Text Node
\draw (191.5,318) node   [align=left] {$\displaystyle x_{3}$};

\end{tikzpicture}

\caption{A distributed encoding system with $K=3$ source nodes, $N=5$ encoding nodes, $\beta=1$ adversarial node, where the adversary is the first source node and sends two different messages to the encoding nodes to confuse the system. The distributed encoding system is required to guarantee the correct decoding of the messages of the honest nodes. For example, if the adversary generates $4$ symbols, and the encoding functions are not designed properly, then we have a set of $5$ equations with $6$ variables, and decoding the messages of the honest nodes might be impossible.
} 
    \label{fig:with adversary}
\end{figure}
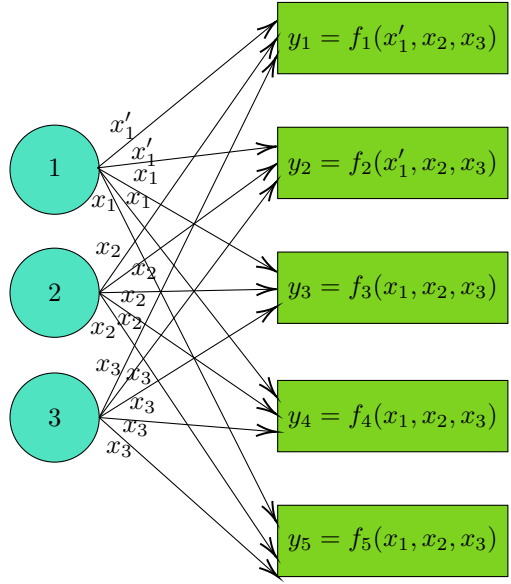

\textbf{Example $1$}. Consider the distributed encoding system depicted in Fig. \ref{fig:with adversary}, wherein $3$ source nodes send messages to $5$ encoding nodes and they store a function of the received messages. In this example, the first source node is the adversary and sends two different messages to the encoding nodes. The second and third source nodes are honest as they send the same message to all encoding nodes. If the first node sends $4$ different symbols to the encoding nodes, then the number of equations, which is $5$, would be less than the number of total variables, which is $6$. In that case, if the encoding functions are not designed properly, then decoding the messages becomes impossible. \\
Suppose the decoder connects to the encoding nodes and has $y_1,y_2,y_3,y_4,y_5$. The decoder does not know the adversary or how it has operated, so it has to take account of all possibilities. In particular, there are $3$ cases for the adversary and $15$ cases for different ways of sending two different messages to the encoding nodes\footnote{There are $5$ cases when the adversary sends $x_1$ to one encoding node and $x'_1$ to the other four encoding nodes. There are $10$ cases when the adversary sends $x_1$ to two encoding nodes and $x'_1$ to the other three encoding nodes. }, amounting to $45$ cases in total. Some cases are shown in table \ref{45 cases}.

\begin{table} \centering \caption{Some of the cases that decoder should consider when decoding the messages of the distributed encoding system of Fig. \ref{fig:with adversary}.}
\begin{tabular}{|l|l|l|l|l|} 
\hline
$f_1(x_1,x_2',x_3)=y_1$ & $f_1(x_1,x_2,x_3)=y_1$  & $f_1(x_1,x_2,x_3)=y_1$  & $f_1(x_1,x_2,x_3)=y_1$  & $f_1(x_1,x_2,x_3)=y_1$  \\
$f_2(x_1,x_2,x_3)=y_2$  & $f_2(x_1,x_2',x_3)=y_2$ & $f_2(x_1,x_2,x_3)=y_2$  & $f_2(x_1,x_2,x_3)=y_2$  & $f_2(x_1,x_2,x_3)=y_2$  \\
$f_3(x_1,x_2,x_3)=y_3$  & $f_3(x_1,x_2,x_3)=y_3$  & $f_3(x_1,x_2',x_3)=y_3$ & $f_3(x_1,x_2,x_3)=y_3$  & $f_3(x_1,x_2,x_3)=y_3$  \\
$f_4(x_1,x_2,x_3)=y_4$  & $f_4(x_1,x_2,x_3)=y_4$  & $f_4(x_1,x_2,x_3)=y_4$  & $f_4(x_1,x_2',x_3)=y_4$ & $f_4(x_1,x_2,x_3)=y_4$  \\
$f_5(x_1,x_2,x_3)=y_5$  & $f_5(x_1,x_2,x_3)=y_5$  & $f_5(x_1,x_2,x_3)=y_5$  & $f_5(x_1,x_2,x_3)=y_5$  & $f_5(x_1,x_2',x_3)=y_5$ \\ \hline
$f_1(x_1,x_2',x_3)=y_1$ & $f_1(x_1,x_2',x_3)=y_1$ & $f_1(x_1,x_2',x_3)=y_1$ & $f_1(x_1,x_2',x_3)=y_1$ & $f_1(x_1,x_2,x_3)=y_1$  \\
$f_2(x_1,x_2',x_3)=y_2$ & $f_2(x_1,x_2,x_3)=y_2$  & $f_2(x_1,x_2,x_3)=y_2$  & $f_2(x_1,x_2,x_3)=y_2$  & $f_2(x_1,x_2',x_3)=y_2$ \\
$f_3(x_1,x_2,x_3)=y_3$  & $f_3(x_1,x_2',x_3)=y_3$ & $f_3(x_1,x_2,x_3)=y_3$  & $f_3(x_1,x_2,x_3)=y_3$  & $f_3(x_1,x_2',x_3)=y_3$ \\
$f_4(x_1,x_2,x_3)=y_4$  & $f_4(x_1,x_2,x_3)=y_4$  & $f_4(x_1,x_2',x_3)=y_4$ & $f_4(x_1,x_2,x_3)=y_4$  & $f_4(x_1,x_2,x_3)=y_4$  \\
$f_5(x_1,x_2,x_3)=y_5$  & $f_5(x_1,x_2,x_3)=y_5$  & $f_5(x_1,x_2,x_3)=y_5$  & $f_5(x_1,x_2',x_3)=y_5$ & $f_5(x_1,x_2,x_3)=y_5$  \\  \hline 
$f_1(x_1,x_2,x_3)=y_1$  & $f_1(x_1,x_2,x_3)=y_1$  & $f_1(x_1,x_2,x_3)=y_1$  & $f_1(x_1,x_2,x_3)=y_1$  & $f_1(x_1,x_2,x_3)=y_1$  \\
$f_2(x_1,x_2',x_3)=y_2$ & $f_2(x_1,x_2',x_3)=y_2$ & $f_2(x_1,x_2,x_3)=y_2$  & $f_2(x_1,x_2,x_3)=y_2$  & $f_2(x_1,x_2,x_3)=y_2$  \\
$f_3(x_1,x_2,x_3)=y_3$  & $f_3(x_1,x_2,x_3)=y_3$  & $f_3(x_1,x_2',x_3)=y_3$ & $f_3(x_1,x_2',x_3)=y_3$ & $f_3(x_1,x_2,x_3)=y_3$  \\
$f_4(x_1,x_2',x_3)=y_4$ & $f_4(x_1,x_2,x_3)=y_4$  & $f_4(x_1,x_2',x_3)=y_4$ & $f_4(x_1,x_2,x_3)=y_4$  & $f_4(x_1,x_2',x_3)=y_4$ \\
$f_5(x_1,x_2,x_3)=y_5$  & $f_5(x_1,x_2',x_3)=y_5$ & $f_5(x_1,x_2,x_3)=y_5$  & $f_5(x_1,x_2',x_3)=y_5$ & $f_5(x_1,x_2',x_3)=y_5$ \\ \hline
\end{tabular} 
\label{45 cases}
\end{table}

\noindent Some applications of the distributed encoding model is as follows.
\begin{itemize}
    \item \emph{Sharding in Blockchain}. One of the main solutions to overcome the scalability issues of blockchains is sharding \cite{shardingFAQ}, which is being studied widely in the blockchain community and incorporated in some blockchain projects such as Zilliqa \cite{zilliqa}, Near \cite{near}, Ethereum \cite{ethereum}, Cardano \cite{cardano}, and PChain \cite{pchain}. In sharding, miners or validators, i.e. processing nodes that can participate in the process of producing or validating blocks, are partitioned into some sets, called shards. Each shard produces its own blocks and has a local chain. The number of validators in a shard is constant, so the number of shards is proportional to the number of validators of the network. This means the rate of block production is proportional to the number of validators, i.e. the network is scalable. One major challenge in sharding is to protect the data of different shards, such that if a fraction of the nodes become out of service, for example, due to DDos attack, the data of all shards can still be recovered. One approach to efficiently resolve this challenge is to encode and store the data of the shards in the nodes \cite{li2018polyshard}.
    
    However, if adversarial validators take over a shard and disperse malicious blocks in the network, the encoded blocks that all the validators of other shards save, would be corrupt. Therefore, considering the distributed nature of data production and also adversarial behavior, the adversarial distributed encoding model fits into this scenario.
    \item \emph{IoT}. In an IoT network, sensors produce data, so the data production process is distributed. As the size of the produced data increases, it becomes impossible to gather the whole data in place, so we need to store them in different places, i.e. in a distributed manner. In order to protect the data, coding is needed before storage. But if the data of some sensors are manipulated adversarially, the stored coded data would be corrupt. Therefore, the adversarial distributed encoding model fits into this scenario as well.
\end{itemize}
\subsection{Related work}
The presence of the adversaries in communication networks and distributed systems is a well studied area. In the context of network coding, a number of papers, as \cite{yeung2006network,cai2006network,jaggi2007resilient} consider a network with an adversary that controls up to a certain number of arbitrary edges with equal capacity and can alter the messages on those edges. In this model, \cite{yeung2006network,cai2006network} provide bounds on the capacity of the network, while \cite{jaggi2007resilient} characterizes the exact capacity and introduces optimal codes. In another model, studied in \cite{kosut2014polytope}, the adversary controls up to a certain number of arbitrary nodes of the network and can put arbitrary messages on the outgoing edges of those nodes. While linear codes are enough to achieve the capacity in the edge-based adversarial model \cite{jaggi2007resilient}, it is shown in \cite{kosut2014polytope} that nonlinear codes are necessary for the node based adversarial model. 

Similar adversarial models are studied in the context of distributed storage systems. In \cite{rashmi2012regenerating}, up to a specific number of erasures and errors occur on the links that connect the storage nodes to a data collector, or the node under repair, during the reconstruction of the data and repair, respectively. \cite{rashmi2012regenerating} has introduced optimal linear coding schemes for the two ``minimum storage regenerating'' and ``minimum bandwidth regenerating'' regions. 
On the other hand, \cite{pawar2011securing,pawar2011securing2,bitar2015securing,kosut2013polytope} consider a distributed storage system wherein the adversary takes over up to a specific number of storage nodes and sends arbitrary data during node repair and data reconstruction. In such model, \cite{pawar2011securing} and \cite{pawar2011securing2} derive upper bounds on the amount of information that can be stored securely in the system, under some conditions, and also provide a coding scheme. \cite{bitar2015securing} extends the results of \cite{pawar2011securing} and \cite{pawar2011securing2} to more general conditions and introduce a capacity achieving scheme. \cite{kosut2013polytope} shows the nonlinear code of \cite{kosut2014polytope} can be used in node-based adversarial distributed storage systems.

Erroneous encoding has been studied in a few works. \cite{yang2014can} and \cite{yang2017computing} study an unreliable encoder made of unreliable gates that each have an error probability. The goal is minimizing the error probability of the encoder with minimum gate redundancy. They find the minimum possible gate redundancy and propose a robust encoder architecture. Authors in \cite{hachem2013coding} consider a faulty linear encoding process, in which the elements of the generator matrix might be erased. They study this model from both probabilistic and worst case approach. In the probabilistic approach, the find an upper bound on the capacity, and in the worst-case approach, find the maximum number of erasures that allows the correct decoding. The same model has been explored in \cite{hosszu2015combinatorial}, where combinatorial methods are used to detect the errors. \cite{dupraz2015evaluation} considers an error probability for each gate in an LDPC encoder and tries to minimize the overall encoding error. \cite{dupraz2016practical} compares the performance of different LDPC encoders in the presence of errors.

\subsection{Contributions}
Our Contributions in this paper are as following.
\begin{enumerate}
    \item Introducing the new problem of distributed encoding (Section \ref{problem formulation}). \\
    The difference between the proposed model and other models in the literature is that in the proposed model, the adversary is in control of a subset of input nodes and their data, and the decoder does not care to decode that part of the data correctly. Indeed, the adversary sacrifices its opportunity to store its own data to mislead the decoder about the data of the honest nodes. However, in all previous models, the input data is intact and the decoder wants to recover all of the input data correctly. Note that the proposed model can not be formulated as a special case of the previous models.
    \item Characterizing the fundamental limit of the distributed \textit{linear} encoding system (Sections \ref{linear achivable section} and \ref{linear converse section}). \\
    In the distributed linear encoding system, encoding nodes use linear functions to encode the data. For this system, we characterize the fundamental limit $t^*_{\textrm{linear}}$ by providing matching achievability and converse bounds. For achievability, we introduce a linear code and show that when the decoder connects to $t^*_{\textrm{linear}}$ or more encoding nodes, it can find the messages of the honest nodes correctly. For the converse, we show that if the decoder connects to less than $t^*_{\textrm{linear}}$ encoding nodes, there is always a way for the adversary to mislead the decoder about the messages of the honest nodes.
    \item Showing that Reed-Solomon codes achieve $t^*_{\textrm{linear}}$ (Section \ref{linear achivable section}).
\end{enumerate}
The paper is organized as follows. We define the problem in Section \ref{problem formulation} and state the main result in Section \ref{main results}. We present achievability and converse proofs for the fundamental limit of the distributed linear encoding system in Sections \ref{linear achivable section} and \ref{linear converse section} respectively. Section \ref{discussion section} is dedicated to discussions and concluding remarks.

\section{Problem Formulation} \label{problem formulation}
We consider an $(N,K,\beta,v)$ \emph{distributed encoding system}, consisting of $K$ source nodes and $N$ encoding nodes, where each source node is directly connected to each of the encoding nodes through a secure and error free link, for some $K,N\in \mathbb{N}$. Let us define $[m]\triangleq \{1,\dots,m\}$, for $m\in \mathbb{N}$. A source node $k\in[K]$ sends a data symbol $x_{n,k}\in \mathbb{F}$ to the encoding node $n\in [N]$, where $\mathbb{F}$ is a large finite field. We use $x_{n,k}$ and $x_{nk}$ interchangeably when it is not confusing. Source nodes are partitioned into two sets $\mathcal{A}$ and $\mathcal{H}$, adversarial and honest source nodes, respectively, where $|\mathcal{A}|\le \beta$, for some $\beta \in \mathbb{N}$, $\beta < K$, and $h\triangleq|\mathcal{H}|$. In other words, there are at most $\beta$ adversarial nodes, and there are $h$ honest nodes. An honest node $k\in \mathcal{H}$ generates only one message $x_k \in \mathbb{F}$ and sends this message to all encoding nodes. Thus, for a source node $k\in \mathcal{H}$, $x_{1k}=x_{2k}=\dots=x_{Nk}=x_k$. However, an adversarial node may generate more than one version for its message to confuse the system. Still, the number of versions that an adversarial node can generate is limited due to some protecting mechanism, such as proof of work.  In particular, we assume that for a source node $k\in \mathcal{A}$, $|\{x_{1k},x_{2k},\dots,x_{Nk}\}|\le v$, for some $v\in \mathbb{N}$. Honest nodes are isolated, and not aware of the messages of the other nodes. However, adversarial nodes are free to collaborate with each other, but still not aware of the messages of the honest nodes. 

The encoding node $n\in [N]$ stores $y_n = f_n(x_{n1},\dots,x_{nK})$, where $f_n:\mathbb{F}^K\rightarrow\mathbb{F}$ is the encoding function of node $n$. The decoding function $g_{\mathcal{T}}:\mathbb{F}^t\rightarrow\mathbb{F}^K$ for a set $\mathcal{T}=\{i_1,\dots,i_t\}\subseteq [N]$, $t\in \mathbb{N}$, is defined as
\begin{align*}
    g_{\mathcal{T}}(y_{i_1},\dots,y_{i_t})=(\hat{x}_1^{\mathcal{T}},\dots,\hat{x}_K^{\mathcal{T}}).
\end{align*}
In other words, $g_{\mathcal{T}}$ takes $y_{i_1},\dots,y_{i_t}$, and produces $(\hat{x}_1^{\mathcal{T}},\dots,\hat{x}_K^{\mathcal{T}})$ as the decoded messages of the source nodes. A code for the distributed encoding system of parameters $(N,K,\beta,v)$, with the $N$ encoding functions $f_1,\dots,f_N$, and the decoding functions $g_{\mathcal{T}}$, for all $\mathcal{T}\subseteq [N]$, $|\mathcal{T}|=t$,  is called \textbf{$\boldsymbol{t}$-correct} if it has the following properties.
 \begin{itemize}
 \item For any $\mathcal{T}\subseteq [N]$, where $|\mathcal{T}|=t$, 
 \begin{align}
    \forall k\in \mathcal{H}: \hat{x}_k^{\mathcal{T}} = x_k \label{correctness condition}.
\end{align}
In other words, the system guarantees correct recovery of the messages of the honest nodes, without necessarily identifying the set of honest and adversarial nodes, or any guarantee to decode the messages of the adversarial nodes.
 \item The encoding is maximum distance separable (MDS), meaning that if all source nodes behave honestly, i.e. $|\{x_{n,k}, n\in [N]\}|=1$ for $k\in [K]$, all $K$ messages can be recovered from any subset of $K$ encoding nodes. 
 \end{itemize}

% A code for the distributed encoding system of parameters $(N,K,\beta,v)$ with $N$ encoding functions $f_1,\dots,f_N$ and decoding functions $g_{\mathcal{T}}$, $\mathcal{T}\subset [N]$, $|\mathcal{T}|=t$,  is  $t$-correct if for any $\mathcal{T}\subset [N]$, where $|\mathcal{T}|=t$ when $0<\beta$ and $|\mathcal{T}|=K$ when $\beta = 0$, the following is satisfied, without having knowledge of $\mathcal{H}$, and the problem formulation does not require decoder to gain such information.
%\begin{align}
%    \forall k\in \mathcal{H}: \hat{x}_k = x_k \label{correctness condition}
%\end{align}
%Note that the condition $|\mathcal{T}|=K$, when $\beta = 0$, is the MDS property when there is no adversary in the system. The system is designed to guarantee correct recovery of the messages of the honest nodes. 

%In a distributed encoding system with a $t$-correct code, we can recover messages of honest source nodes correctly. 
Whenever it is not confusing, we denote $\hat{x}_k^{\mathcal{T}}$ by $x_k$.
For an $(N,K,\beta,v)$ distributed encoding system, we define $t^*(N,K,\beta,v)$ as the minimum $t$ for which a $t$-correct code exists.
 If we restrict the encoding functions  $f_1,\dots,f_N$ to be linear functions, we denote this notion by $t^*_{\textrm{linear}}(N,K,\beta,v)$. 
 In this paper, the objective is to characterize  $t^*_{\textrm{linear}}(N,K,\beta,v)$.

\section{Main Result} \label{main results}
The following theorem states the main result of this paper for an $(N,K,\beta,v)$ distributed encoding system.

\begin{theorem}\label{linear theorem}
In an $(N,K,\beta,v)$ distributed encoding system,
\begin{align}
    t^*_{\textrm{linear}}(N,K,\beta,v) = \begin{cases}
       K+2\beta(v-1)  &  \mbox{if $N\geq K+2\beta(v-1)$,}\\
       N  & \mbox{if $K\le N< K+2\beta(v-1)$.}
    \end{cases}
\end{align}
\end{theorem}
 The achievability proof for this theorem is in Section \ref{linear achivable section} and the converse proof is in Section \ref{linear converse section}.
 In the case $N\ge K+2\beta(v-1)$, $t^*_{\textrm{linear}}(N,K,\beta,v)=K+2\beta(v-1)$, i.e. $t^*_{\textrm{linear}}$ is independent from $N$. Note that although there are at most $K+\beta(v-1)$ variables in an $(N,K,\beta,v)$ distributed encoding system (adversarial nodes have $\beta(v-1)$ excessive messages), we need at least $K+2\beta(v-1)$ linear equations to produce $(\hat{x}_1,\dots,\hat{x}_K)$ where $\forall k\in \mathcal{H}, \hat{x}_k=x_k$. That is to say, $\beta(v-1)$ extra equations are needed. The reason for this redundancy in the number of equations is that the adversarial source nodes can choose their messages in a such way that \emph{shifts} the messages of the honest nodes, without knowing them. In other words, the adversary can exploit the linearity of the equations to cause the decoder to decode $\hat{x}_k=x_k+\Delta_k$ instead of $x_k$, for at least one $k\in \mathcal{H}$. Let us emphasize that we do not simply have a linear set of equations to solve for the data. Rather, the decoder needs to consider various ways that the adversary can use its options, and see if each one leads to a feasible solution. Therefore, inherently, the conventional limits for a system of linear equations do not apply here. We emphasize that identifying the set of adversarial nodes or decoding the messages of adversarial nodes correctly is not necessary to satisfy the correctness condition \eqref{correctness condition}.

\section{Achievability proof for $t^*_{\textrm{linear}}(N,K,\beta,v)$}\label{linear achivable section}
First consider the simple case $K\le N\le K+2\beta( v-1)$. For this case, Theorem \ref{linear theorem} states $t^*_{\textrm{linear}}=N$. The proposed $N$-correct linear code for this case is
\begin{align}
    f_n(x_{n1},\dots, x_{nK})  &= x_{nn}, \quad n\in[K], \label{code for k<=n<=h+2av-a} \\
    f_n(x_{n1},\dots, x_{nK})  &=  L_n(x_{n1}, \ldots,   x_{nK}), \quad K+1\leq n \leq N, \label{remainign f linear}
\end{align}
where $L_n:\mathbb{F}^K\rightarrow \mathbb{F}$ denotes a linear function whose coefficients are chosen independently and uniformly at random from $\mathbb{F}$.
Since $t^*_{\textrm{linear}}=N$ in this case, $\mathcal{T}=[N]$, and the decoder has all the $N$ coded values, including those in \eqref{code for k<=n<=h+2av-a}, so it can declare $(\hat{x}_1,\dots,\hat{x}_K)= (x_{11},\dots,x_{KK})$. For node $k\in \mathcal{H}$, $x_{kk}=x_k$, so the correctness condition  \eqref{correctness condition} is satisfied. The MDS property is also satisfied with high probability due to the randomness of the coefficients in \eqref{remainign f linear}. For $N\ge K+2\beta(v-1)$, we parse the achievability proof into three parts: code construction, decoding algorithm, and proof of correctness. 
  \subsection{Code Construction}
  we propose the following linear code.
\begin{align} \label{our linear code}
    f_n(x_{n1},\dots,x_{nK})= L_n(x_{n1},\dots,x_{nK}), \quad n\in[N].
\end{align}
Again, due to the random coefficients of $L_1,\dots,L_N$, this code has MDS property with high probability. The decoding procedure for this code is described in Algorithm \ref{decoding alg}, for $t=t^*_{\textrm{linear}}=K+2\beta (v-1) = h+2\beta v-\beta$. 
 
\begin{algorithm} 
\caption{Decoding Algorithm}
\label{decoding alg}
\SetAlgoLined
\KwIn{$\mathcal{T}\subseteq[N], |\mathcal{T}|=t, \{y_n\}_{n\in \mathcal{T}}$}
\KwOut{$\hat{x}_1,\dots,\hat{x}_K$}
\For{$k\in [K]$}{$\hat{x}_k \gets \perp$} \
\For{$\mathcal{\hat{A}}\in[K], |\mathcal{\hat{A}}|=\beta$ \label{loop begin}}{
$\hat{\mathcal{H}}=[K]\setminus \mathcal{\hat{A}}$\;
\For{$k \in \mathcal{\hat{H}}$}{
\For{$n\in \mathcal{T}$}{
$x_{nk}=\tilde{x}_k$}}
\For{$\big\{\{\mathcal{I}_k^{(1)},\dots,\mathcal{I}_k^{(v)}\},k\in\mathcal{\hat{A}}\big\},$ \textbf{such that for all} $k\in\mathcal{\hat{A}}$\textbf{, we have} $\mathcal{I}_k^{(1)}\cup \dots \cup \mathcal{I}_k^{(v)}=\mathcal{T}, \mathcal{I}_k^{(i)}\cap\mathcal{I}^{(j)}_k=\varnothing, 1\le i< j\le v$\label{partitioning}}{ 
\For{$k\in \mathcal{\hat{A}}$}{
\For{$i\in[v]$}{
\For{$n\in \mathcal{I}^{(i)}_k$}{
$x_{nk} = \tilde{x}^{(i)}_k$ \label{i th version}}
}}
\If{the system of equations $f_n(x_{n1},\dots,x_{nK})=y_n, n\in \mathcal{T}$ has a solution for $\{\tilde{x}_k\}_{k\in \mathcal{\hat{H}}}$ and $\{\tilde{x}_k^{(i)}\}_{k\in \mathcal{\hat{A}},i\in[v]}$ \label{system of equations}}{
 \For{$k \in \mathcal{\hat{H}}$}{\If{$\hat{x}_k=\perp$}{
$\hat{x}_k = \tilde{x}_k$ \label{save honest messages}}}}\label{end last if} 
\label{loop end}} \label{end of partitioning}} \label{algo end}
\end{algorithm}

\subsection{Decoding Algorithm}
In the first step of the decoding algorithm, the decoder assumes $\beta$ source nodes as the adversaries, called the \textit{presumed} adversarial nodes, and denotes the set by $\mathcal{\hat{A}}$. Similarly, $\mathcal{\hat{H}}$ is the set of the presumed honest nodes. Not knowing the real adversaries $\mathcal{A}$, the decoder examines all the cases of $\mathcal{\hat{A}}$. This corresponds to the loop in lines \ref{loop begin} to \ref{algo end} in Algorithm \ref{decoding alg}.

In the second step, for each presumed adversarial node, the decoder considers a partitioning of encoding nodes based on which that adversarial node sends its messages. In other words, that partitioning specifies the encoding nodes that receive the same message from that node. Let $\mathcal{I}^{(i)}_k\subseteq[N]$ be the set of the encoding nodes that receive $\tilde{x}_k^{(i)}$, $i\in[v]$, from the source node $k\in\mathcal{\hat{A}}$. All the encoding nodes receive $\tilde{x}_k$ from $k\in\mathcal{\hat{H}}$. All the variables are initialized with $\perp$. The decoder examines all the possible partitionings for all $k\in\mathcal{\hat{A}}$, which corresponds to lines \ref{partitioning} to \ref{end of partitioning} in Algorithm \ref{decoding alg}.

In the third step, for each choice of $\mathcal{\hat{A}}$ and $\mathcal{I}^{(1)}_k,\dots, \mathcal{I}^{(v)}_k$, $k\in\mathcal{\hat{A}}$, the decoder forms a linear system of $t^*_{\textrm{linear}}=h+2\beta v-\beta$ equations with the $h+\beta v$ variables $\{\tilde{x}_k, k\in \mathcal{\hat{H}}\}$ and $\{\tilde{x}_k^{(i)}, k\in \mathcal{\hat{A}},i\in[v]\}$.
The decoder solves this system of equations and if a solution exists, stores only the solution for the messages of the presumed honest nodes. Since there are more equations than variables, not all cases result in solutions. This step corresponds to lines \ref{system of equations} to \ref{end last if} in Algorithm \ref{decoding alg}.

In the decoding algorithm, whenever the decoder finds a feasible solution, it only saves the solution of the presumed honest nodes, if a previous value has not been saved for that variable. The reason for this strategy is three-fold:
\begin{enumerate}
    \item
    One of the feasible solutions that the decoder finds, corresponds to the real scenario. Therefore, the messages of the honest nodes do not remain unassigned after the termination of the algorithm.
    \item 
    In the next part, correctness proof, we prove that if the decoder begins with an $\mathcal{\hat{H}}$, and finds a feasible solution, the value of the message of $k\in\mathcal{H}\cap\mathcal{\hat{H}}$ in the solution is correct.
    \item
    There is no need for the correctness of the messages of the adversarial node. Thus, any value for the message of the source node $k\in\mathcal{\hat{H}}, k\not\in\mathcal{H}$, does not affect the correctness of the algorithm.
\end{enumerate}
Therefore, when the decoder assigns a value to the message of an honest node, we are assured about the correctness of that value. It is important to note that the output of the decoding algorithm, $\hat{x}_1,\dots,\hat{x}_K$, does not necessarily satisfy the equations $f_n(\hat{x}_1,\dots,\hat{x}_K)=y_n, n\in[N]$. In fact, some of $\hat{x}_1,\dots,\hat{x}_K$ might even remain as $\perp$. Nevertheless, the correctness condition is satisfied.

\subsection{Correctness Proof}
In the following lemma, we show that the proposed linear code satisfies the correctness condition in Lemma \ref{linear uniqueness}.

\begin{lemma} \label{linear uniqueness}
The linear code described in \eqref{our linear code}, with the decoding in the Algorithm \ref{decoding alg} and $t\ge t^*_{\textrm{linear}}=h+2\beta v-\beta$, satisfies the correctness condition \eqref{correctness condition}.
\end{lemma}

Before proving this lemma, we need the following lemma.

\begin{lemma} \label{2v-1 lemma}
Consider the sets $\mathcal{A}=\{a_1,\dots,a_v\}\subseteq \mathbb{F}^v$ and $\mathcal{B}=\{b_1,\dots,b_v\}\subseteq \mathbb{F}^v$, and $\mathcal{C}=\{c_{i,j}=a_i-b_j, i,j\in[v]\}$. There exists a subset $\mathcal{C'}=\{c_{i_1},\dots,c_{i_{2v-1}}\}\subseteq \mathcal{C}$, $|\mathcal{C'}|=2v-1$, such that 
each element of $\mathcal{C}$ is a linear combination of elements of $\mathcal{C'}$, with constant coefficients, i.e. not functions of $a_i, b_j$, $i,j\in[v]$.
\end{lemma}
As an example, consider $\mathcal{A}=\{a_1,a_2\}$, $\mathcal{B}=\{b_1,b_2\}$. So $\mathcal{C}=\{a_1-b_1,a_1-b_2,a_2-b_1,a_2-b_2\}$.
Note that
\begin{align*}
    a_2-b_2=(a_2-b_1)-(a_1-b_1)+(a_1-b_2).
\end{align*}
Therefore $\mathcal{C'}=\{a_1-b_1,a_1-b_2,a_2-b_1\}$ satisfies the lemma.
\begin{proof}
We show that $\mathcal{C}' = \{c_{i,i}, i\in[v]\} \cup \{c_{i,i+1}, i\in[v-1]\}$, $|\mathcal{C'}|=2v-1$, satisfies the lemma. Consider any $c_{i,j}\in \mathcal{C}$.
\begin{itemize}
\item If $j \ge i+2$:
\begin{align} \hspace*{-0.5cm}
c_{ij} = a_i - b_j = & \big[ (a_i- b_{i+1})+\dots+(a_{j-1}-b_j)\big] -  \nonumber \\ & \big[ (a_{i+1}-b_{i+1})+\dots+(a_{j-1} - b_{j-1})\big] \nonumber \\
= & (c_{i,i+1}+\dots+c_{j-1,j}) - (c_{i+1,i+1}+\dots+c_{j-1,j-1})
\end{align}
        
\item
If $j \le i-1$:
\begin{align}
c_{ij} = a_i - b_j = & \big[ (a_i-b_i)+\dots+(a_j-b_j)\big] -  \nonumber \\ & \big[ (a_{i-1}-b_i)+\dots+(a_{j}-b_{j+1})\big] \nonumber \\
= & (c_{i,i}+\dots+c_{j,j}) - (c_{i-1,i}+\dots+c_{j,j+1})
\end{align}
\end{itemize}
Another choice for $\mathcal{C}'$ is $\mathcal{C'}=\{c_{1,i}, i\in[v]\} \cup \{c_{i,i}, 2\le i\le v\}$. Then, for any $c_{i,j}\in \mathcal{C}$,
\begin{align}
c_{ij} = a_i - b_j = (a_i-b_i)-(a_1-b_i)+(a_1-b_j) = c_{i,i}-c_{1,i}+c_{1,j}.
\end{align}
Consequently, there is more than one choice for $\mathcal{C}'$.
\end{proof}

Now we prove Lemma \ref{linear uniqueness}.

\begin{proof}[Proof of Lemma \ref{linear uniqueness}]
Without loss of generality, suppose that the first $\beta$ source nodes are indeed the adversaries, i.e. $\mathcal{A}=\{1,\dots, \beta\}$ and $\mathcal{H}=\{\beta+1,\dots,K\}$, and of course the decoder is not aware of $\mathcal{A}$ and $\mathcal{H}$. We denote the $v$ messages of the adversarial source node $k\in \mathcal{A}$ by $\{x_k^{(i)}\}_{i\in[v]}$, $k\in[\beta]$, and the message of the honest node $k\in \mathcal{H}$ by $x_k$, $\beta+1\le k\le K$. Assuming an arbitrary set $\mathcal{T}=\{n_1,\dots,n_t\}$, we want to show that the only solution for the messages of the honest nodes in  $f_n(x_{n1},\dots,x_{nK})=y_n$, $n\in \mathcal{T}$, is $x_{nk}=x_k$, $k\in\mathcal{H}$. We prove Lemma \ref{linear uniqueness} for $t=t^*_{Linear}$. The proof for $t>t^*_{Linear}$ is trivially followed. Suppose that in addition to the real scenario, the decoder finds another solution for $f_n(x_{n1},\dots,x_{nK})=y_n$, $n\in \mathcal{T}$, with $\mathcal{\hat{A}}$ as the presumed adversarial nodes, $\{z_k^{(i)}\}_{i\in[v]}$ as the messages of the node $k\in\mathcal{\hat{A}}$, $\mathcal{\hat{H}}$ as the presumed honest nodes, and $z_k$ as the message of the node $k\in\mathcal{\hat{H}}$, where $\exists k\in \mathcal{\hat{H}}\cap\mathcal{H}: z_k\neq x_k$. In the following, we consider all possible cases for $\mathcal{\hat{A}}$, and show that this is indeed impossible. There are three possible cases, $\mathcal{\hat{A}}=\mathcal{A}$, and $\mathcal{\hat{A}}\neq\mathcal{A}, \mathcal{\hat{A}}\cap\mathcal{A}\neq\varnothing$, and $\mathcal{\hat{A}}\cap\mathcal{A}=\varnothing$. 
Let $z_k=x_k+\Delta_k$ for $k\in \mathcal{\hat{H}}\cap \mathcal{H}$. The goal is to prove $\Delta_k=0$ for $k\in \mathcal{\hat{H}}\cap \mathcal{H}$.\\
\large \textbf{Case$\mathbf{1}$. $\mathbf{\mathcal{\hat{A}}=\mathcal{A}}$}. \\ \normalsize
\textbf{Step $\mathbf{1}$}. We have the following linear system of equations.
\begin{align} \label{first lienar}
    L_n(z_{n1},\dots,z_{n\beta},x_{\beta+1}+\Delta_{\beta+1},\dots,x_K+\Delta_K) = L_n(x_{n1},\dots,x_{n\beta},x_{\beta+1},\dots,x_K), \quad n\in \mathcal{T},
\end{align}
where $\{z_{nk}\}_{n\in \mathcal{T}}=\{z_k^{(i)}\}_{i\in[v]}$, and $\{x_{nk}\}_{n\in \mathcal{T}}=\{x_k^{(i)}\}_{i\in[v]}$, $k\in[\beta]$. In the following, we show that we can rewrite the above equations with $h+2\beta v-\beta$ variables. In addition, those equations are linearly independent and the adversary cannot make them dependent by exploiting its options.\\
The above is equivalent to
\begin{align}
L_n(z_{n1}-x_{n1},\dots,z_{n\beta}-x_{n\beta},\Delta_{\beta+1},\dots,\Delta_K) = 0, \quad n\in \mathcal{T}. \label{the new linear system}
\end{align}
This linear system of equations can be expressed as
\begin{align}
\sum_{k\in[\beta]}\mathbf{B}_k
\begin{bmatrix}
z_{n_1k}-x_{n_1k} \\
\vdots\\
z_{n_tk}-x_{n_tk}
\end{bmatrix}
+ \mathbf{C}
\begin{bmatrix}
\Delta_{\beta+1} \\
\vdots \\
\Delta_K
\end{bmatrix}
= \begin{bmatrix}
0 \\
\vdots \\
0
\end{bmatrix}, \label{matrix translation}
\end{align}
where $\mathbf{B}_k$, $k\in[\beta]$, and $\mathbf{C}$ are $t\times t$ and $t\times (K-\beta)$ matrices respectively, and are composed of the coefficients of the linear code. Let the coefficients of the proposed linear code be $\alpha_{nk}$, $n\in [N], k\in [K]$, i.e. $f_n(x_{n1},\dots,x_{nK})=\sum_{k=1}^K \alpha_{nk}x_{nk}$. 
Thus, 
\begin{align}
    \mathbf{B}_k = \begin{bmatrix}
    \alpha_{n_1,k} & 0 & \dots & 0 \\
    0 & \alpha_{n_2,k} & \dots & 0 \\
     &  & \ddots &  \\
    0 & 0 & \dots & \alpha_{n_t,k} \\
    \end{bmatrix}, \quad k\in[\beta],
\end{align}
and
\begin{align}
    \mathbf{C} = \begin{bmatrix}
    \alpha_{n_1,\beta+1} & \dots &  \alpha_{n_1,K} \\
     & \ddots & \\
     \alpha_{n_t,\beta+1} & \dots &  \alpha_{n_t,K}
    \end{bmatrix}.
\end{align}
\textbf{Step $\mathbf{2}$}. We know that
\begin{align*}
    \{z_{nk}-x_{nk}\}_{1\le n\le t}\subseteq\{z_k^{(i)}-x_k^{(j)}\}_{i,j\in[v]}, \quad k\in[\beta].
\end{align*}
According to Lemma \ref{2v-1 lemma}, there exist 
\begin{align*}
\{w_{1,k},\dots,w_{2v-1,k}\}\subseteq \{z_k^{(i)}-x_k^{(j)}\}_{i,j\in[v]},
\end{align*}
and $t\times (2v-1)$ constant tall\footnote{By a ``tall'' matrix, we mean a matrix with more rows than columns. By a ``fat'' matrix, we mean the opposite.} matrices $\mathbf{D}_k$, such that
\begin{align} \label{first case D_k}
    \begin{bmatrix}
z_{n_1k}-x_{n_1k} \\
\vdots\\
z_{n_tk}-x_{n_tk}
\end{bmatrix} = \mathbf{D}_k \begin{bmatrix}
w_{1,k} \\
\vdots \\
w_{2v-1,k}
\end{bmatrix}, \quad k\in[\beta],
\end{align}
where $\mathbf{D}_k$ is comprised of $0,\pm 1$ elements. Now, \eqref{matrix translation} can be rewritten as
\begin{align}
\sum_{k\in[\beta]}\mathbf{B}_k\mathbf{D}_k
\begin{bmatrix}
w_{1,k} \\
\vdots \\
w_{2v-1,k}
\end{bmatrix}
+ \mathbf{C}
\begin{bmatrix}
\Delta_{\beta+1} \\
\vdots \\
\Delta_K
\end{bmatrix}
= \begin{bmatrix}
0 \\
\vdots \\
0
\end{bmatrix}. \label{matrix translation, 2v-1}
\end{align}
The element in row $n_i$, $i\in[t]$ of $\mathbf{B}_k\mathbf{D}_k$ are either $0$, $\alpha_{n_i,k}$, or $-\alpha_{n_i,k}$. The linear system of equations in \eqref{matrix translation, 2v-1} has $\beta(2v-1)+h$ variables and $t=\beta(2v-1)+h$ equations. Recall that we want to show $\Delta_{\beta+1}=\dots=\Delta_K=0$. Let $\mathbf{B}\triangleq[\mathbf{B}_1\mathbf{D}_1 | \dots|\mathbf{B}_{\beta}\mathbf{D}_{\beta}]$, and $\textrm{rank}(\mathbf{B})=r$ for some $r\in \mathbb{N}$. Since $\mathbf{B}$ is a tall $t\times (2\beta v-\beta)$ matrix, we have $r\le 2\beta v-\beta$. For a full rank submatrix of $\mathbf{B}$ of size $t\times r$, containing the $r$ linearly independent columns of $\mathbf{B}$, which we denote by $\mathbf{\tilde{B}}$, there exists a variable vector $[w'_1,\dots,w'_r]$, such that
\begin{align} \label{A tilde}
\sum_{k\in[\beta]}\mathbf{B}_k\mathbf{D}_k
\begin{bmatrix}
w_{1,k} \\
\vdots \\
w_{2v-1,k}
\end{bmatrix} = \mathbf{\tilde{B}} \begin{bmatrix}
w'_1 \\
\vdots \\
w'_r
\end{bmatrix}.
\end{align}
Without loss of generality, suppose that $\mathbf{\tilde{B}}$ is the first $r$ columns of $\mathbf{B}$. Similar to $\mathbf{B}$, each elements in row $n_i$, $i\in[t]$ of $\mathbf{\tilde{B}}$ belongs to $\{0,\pm\alpha_{n_i,1},\dots,\pm\alpha_{n_i,\beta}\}$. Thus, we can denote $\mathbf{\tilde{B}}$ by
\begin{align}
    \mathbf{\tilde{B}} = \begin{bmatrix}
    \gamma_{n_1,1}\alpha_{n_1,1} & \dots & \gamma_{n_1,r}\alpha_{n_1,r} \\
    & \ddots & \\
    \gamma_{n_t,1}\alpha_{n_t,1} & \dots & \gamma_{n_t,r}\alpha_{n_t,r} 
    \end{bmatrix},
\end{align}
where $\gamma_{n_i,j}\in \{0,1,-1\}$, $i\in [t], j\in [r]$. The coefficients $\gamma_{n_i,j}$'s are determined by $\mathbf{D}_1,\dots,\mathbf{D}_{\beta}$, and therefore are dependent on the way the adversary uses its options on $z_{n_1 k},\dots,z_{n_t k}$ and $x_{n_1 k},\dots,x_{n_t k}$, according to \eqref{first case D_k}. Thus, $\gamma_{n_i,j}$'s are in control of the adversary.
Using \eqref{A tilde}, we rewrite \eqref{matrix translation, 2v-1} as
\begin{align} \label{final linear}
[\mathbf{\tilde{B}}|\mathbf{C}]\begin{bmatrix}
w'_1 \\
\vdots \\
w'_r \\
\Delta_{\beta+1} \\
\vdots \\
\Delta_K
\end{bmatrix} = \begin{bmatrix}
0\\
\vdots\\
0
\end{bmatrix}.
\end{align}
\textbf{Step $\mathbf{3}$}. We show that $[\mathbf{\tilde{B}}|\mathbf{C}]$ is full rank. For this, first we show that the concatenation of $\mathbf{\tilde{B}}$ and the first column of $\mathbf{C}$, i.e.
\begin{align}
     \begin{bmatrix}
    \gamma_{n_1,1}\alpha_{n_1,1} & \dots & \gamma_{n_1,r}\alpha_{n_1,r} &  \alpha_{n_1,\beta+1}\\
    & & \ddots & \\
    \gamma_{n_t,1}\alpha_{n_t,1} & \dots & \gamma_{n_t,r}\alpha_{n_t,r} & \alpha_{n_t,\beta+1}
    \end{bmatrix}, \label{induction level 1}
\end{align}
is full rank. Since $\mathbf{\tilde{B}}$ is a $t\times r$ tall full rank matrix, without loss of generality, we assume that the first $r$ rows in it are linearly independent. By contradiction, suppose that the last column of \eqref{induction level 1} is a linear combination of the other $r$ columns, i.e.
\begin{align}
    \zeta_1 \begin{bmatrix}
    \gamma_{n_1,1}\alpha_{n_1,1} \\
    \vdots \\
    \gamma_{n_t,1}\alpha_{n_t,1}
    \end{bmatrix} + \dots + \zeta_r \begin{bmatrix}
    \gamma_{n_1,r}\alpha_{n_1,r} \\
    \vdots \\
    \gamma_{n_t,r} \alpha_{n_t,r}
    \end{bmatrix} = \begin{bmatrix}
    \alpha_{n_1,\beta+1} \\
    \vdots \\
    \alpha_{n_t,\beta+1}
    \end{bmatrix}, \label{zeta linear combination}
\end{align}
where $\zeta_1,\dots,\zeta_r\in \mathbb{F}$. Since the first $r$ rows of $\mathbf{\tilde{B}}$ constitute a full rank square matrix, $\zeta_1,\dots,\zeta_r$ are uniquely found by
\begin{align}
    \begin{bmatrix}
    \zeta_1 \\
    \vdots \\
    \zeta_r
    \end{bmatrix} = \begin{bmatrix}
    \gamma_{n_1,1}\alpha_{n_1,1} & \dots & \gamma_{n_1,r}\alpha_{n_1,r} \\
    & \ddots & \\
    \gamma_{n_r,1}\alpha_{n_r,1} & \dots & \gamma_{n_r,r}\alpha_{n_r,r} 
    \end{bmatrix}^{-1} \begin{bmatrix}
    \alpha_{n_1,\beta+1} \\
    \vdots \\
    \alpha_{n_r,\beta+1}
    \end{bmatrix}. \label{unique zeta}
\end{align}
But $r+1\le 2\beta v-\beta+1\le t$, so $\zeta_1,\dots,\zeta_r$ must satisfy the equations of the remaining rows in \eqref{induction level 1}, the first of which is
\begin{align}
   \zeta_1\gamma_{n_{r+1},1}\alpha_{n_{r+1},1}+\dots + \zeta_r\gamma_{n_{r+1},r}\alpha_{n_{r+1},r} = \alpha_{n_{r+1},\beta+1}. \label{linear zeta}
\end{align}
We know that the only leverage of the adversary in \eqref{unique zeta} and \eqref{linear zeta} is  $\gamma_{n_{i},j}, i\in[r+1],j\in[r]$, but  $\gamma_{n_{i},j}\in\{0,1,-1\}$. So, the adversary has a limited number of trials\footnote{Note that all $3^{r(r+1)}$ values for $\gamma_{n_{i},j}, i\in[r+1],j\in[r]$ are not possible. For example all $0$ or all $1$ cases are not possible.}. Since $\alpha_{n_{r+1},\beta+1}$ is chosen  uniformly at random from a large field, and independently from $\{\alpha_{n_i,j}, i\in[r+1],j\in[r]\}$ and $\{\alpha_{n_i,\beta+1},i\in[r]\}$, the probability that its choice is such that the adversary can satisfy \eqref{linear zeta} by choosing $\{\gamma_{n_i,j},i\in[r+1],j\in[r]\}$, is negligible. Thus, we conclude that the matrix in \eqref{induction level 1} is full rank with high probability. Similarly, we can prove that
\begin{align}
     \begin{bmatrix}
    \gamma_{n_1,1}\alpha_{n_1,1} & \dots & \gamma_{n_1,r}\alpha_{n_1,r} &  \alpha_{n_1,\beta+1} & \alpha_{n_1,\beta+2}\\
    & & \ddots & & \\
    \gamma_{n_t,1}\alpha_{n_t,1} & \dots & \gamma_{n_t,r}\alpha_{n_t,r} & \alpha_{n_t,\beta+1} & \alpha_{n_t,\beta+2}
    \end{bmatrix}, \label{induction level 2}
\end{align}
and finally
\begin{align}
    [\mathbf{\tilde{B}}|\mathbf{C}] = \begin{bmatrix}
    \gamma_{n_1,1}\alpha_{n_1,1} & \dots & \gamma_{n_1,r}\alpha_{n_1,r} &  \alpha_{n_1,\beta+1} & \dots & \alpha_{n_1,\beta+h}\\
    & & \ddots & & & \\
    \gamma_{n_t,1}\alpha_{n_t,1} & \dots & \gamma_{n_t,r}\alpha_{n_t,r} & \alpha_{n_t,\beta+1} & \dots & \alpha_{n_t,\beta+h}
    \end{bmatrix}, \label{induction level h}
\end{align}
are full rank.\\
\textbf{Step $\mathbf{4}$}. Now that $[\mathbf{\tilde{B}}|\mathbf{C}]$ is a full rank matrix of size $t\times (r+h)$, and $r+h\le 2\beta v\beta +h= t$, the unique solution of \eqref{final linear} is
\begin{align*}
    w'_1=\dots=w'_r=\Delta_{\beta+1}=\dots=\Delta_K=0.
\end{align*}\\
\large \textbf{Case$\mathbf{2}$. $\mathbf{\mathcal{\hat{A}}\neq \mathcal{A}, \mathcal{\hat{A}}\cap\mathcal{A}\neq\varnothing}$.}\\ \normalsize
Without loss of generality, we assume that $\mathcal{\hat{A}}=\{m,\dots,\beta+m-1\}$ for some $2\le m\le \textrm{min}(\beta,h)$. \\
\textbf{Step $\mathbf{1}$}. We have the following linear system of equations.
\begin{align*}
    L_n(z_1,\dots,z_{m-1},z_{n,m},\dots,z_{n,\beta+m-1},x_{\beta+m}+\Delta_{\beta+m},\dots,x_K+\Delta_K) =  L_n(x_{n1},\dots,x_{n\beta},x_{\beta+1},\dots,x_K), n\in \mathcal{T},
\end{align*}
where $\{z_{n,k}\}_{n\in \mathcal{T}}=\{z_k^{(i)}\}_{i\in[v]}$ for $k\in \mathcal{\hat{A}}$, and $\{x_{n,k}\}_{n\in \mathcal{T}}=\{x_k^{(i)}\}_{i\in[v]}$ for $k\in[\beta]$. In the following, we show that we can rewrite the above equations with $h+2\beta v-\beta$ variables. In addition, those equations are linearly independent and the adversary cannot make them dependent by exploiting its options.\\
The above system of equations is equivalent to
\begin{align}
L_n(z_1-x_{n1},\dots,z_{m-1}-x_{n,m-1}, z_{n,m}-x_{n,m},\dots,z_{n,\beta}-x_{n,\beta},z_{n,\beta+1}&-x_{\beta+1},\dots,z_{n,\beta+m-1}-x_{\beta+m-1}, \nonumber \\
& \Delta_{\beta+m},\dots,\Delta_K) = 0, \quad n\in \mathcal{T}. \label{the new linear system, second case}
\end{align}
This system of equations can be expressed as
\begin{align}
\sum_{k=1}^{m-1}\mathbf{B}_k \begin{bmatrix}
z_k-x_{n_1,k} \\
\vdots \\
z_k-x_{n_t,k}
\end{bmatrix} +
\sum_{k=m}^{\beta}\mathbf{B}_k
\begin{bmatrix}
z_{n_1,k}-x_{n_1,k} \\
\vdots\\
z_{n_t,k}-x_{n_t,k}
\end{bmatrix} + \sum_{k=\beta+1}^{\beta+m-1}\mathbf{B}_k \begin{bmatrix}
z_{n_1,k}-x_k \\
\vdots \\
z_{n_t,k}-x_k
\end{bmatrix} + \mathbf{C}
\begin{bmatrix}
\Delta_{\beta+m} \\
\vdots \\
\Delta_K
\end{bmatrix}
= \begin{bmatrix}
0 \\
\vdots \\
0
\end{bmatrix}, \label{matrix translation, second case}
\end{align}
where $\mathbf{B}_k$, $k\in[\beta+m-1]$, and $\mathbf{C}$ are $t\times t$ and $t\times (h-m+1)$ matrices respectively, and are composed of the coefficients of the linear code. \\
\textbf{Step $\mathbf{2}$}. We know that
\begin{align*}
    \{z_{nk}-x_{nk}\}_{n\in \mathcal{T}}\subseteq\{z_k^{(i)}-x_k^{(j)}\}_{1\le i,j\le v}, \quad m\le k\le \beta.
\end{align*}
According to Lemma \ref{2v-1 lemma}, there exist 
\begin{align*}
\{w_{1,k},\dots,w_{2v-1,k}\}\subseteq \{z_k^{(i)}-x_k^{(j)}\}_{1\le i,j\le v}, \quad m\le k\le \beta,
\end{align*}
and $t\times(2v-1)$ constant matrices, $\mathbf{D}_k$, such that
\begin{align}
\begin{bmatrix}
z_{n_1k}-x_{n_1k} \\
\vdots\\
z_{n_tk}-x_{n_tk}
\end{bmatrix} = \mathbf{D}_k \begin{bmatrix}
w_{1,k} \\
\vdots \\
w_{2v-1,k}
\end{bmatrix}, \quad m\le k\le \beta.
\end{align}
Furthermore, since \[\{z_k-x_{n,k}\}_{n\in \mathcal{T}}=\{z_k-x_k^{(i)}\}_{i\in[v]}\triangleq \{w_{i,k}\}_{i\in[v]},\quad 1\le k\le m-1\] and 
\begin{align*}
\{z_{n,k}-x_k\}_{n\in \mathcal{T}} =  \{z_k^{(i)}-x_k\}_{i\in[v]} \triangleq \{w_{i,k}\}_{i\in[v]}, \quad \beta+1\le k\le \beta+m-1
\end{align*}
there exist $t\times v$ constant matrices $\mathbf{E}_k$, such that 
\begin{align}
\begin{bmatrix}
z_k-x_{n_1,k} \\
\vdots\\
z_k-x_{n_t,k}
\end{bmatrix} = \mathbf{E}_k \begin{bmatrix}
w_{1,k} \\
\vdots \\
w_{v,k}
\end{bmatrix}, \quad 1\le k\le m-1
\end{align}
and 
\begin{align}
\begin{bmatrix}
z_{n_1,k}-x_k \\
\vdots\\
z_{n_t,k}-x_k
\end{bmatrix} = \mathbf{E}_k \begin{bmatrix}
w_{1,k} \\
\vdots \\
w_{v,k}
\end{bmatrix}, \quad \beta+1\le k\le \beta+m-1.
\end{align}
Therefore, \eqref{matrix translation, second case} can be rewritten as
\begin{align}
\sum_{k=1}^{m-1}\mathbf{B}_k\mathbf{E}_k \begin{bmatrix}
w_{1,k} \\
\vdots \\
w_{v,k}
\end{bmatrix} +
\sum_{k=m}^{\beta}\mathbf{B}_k\mathbf{D}_k
\begin{bmatrix}
w_{1,k} \\
\vdots\\
w_{2v-1,k}
\end{bmatrix} + \sum_{k=\beta+1}^{\beta+m-1}\mathbf{B}_k\mathbf{E}_k \begin{bmatrix}
w_{1,k} \\
\vdots \\
w_{v,k}
\end{bmatrix} + \mathbf{C}
\begin{bmatrix}
\Delta_{\beta+m} \\
\vdots \\
\Delta_K
\end{bmatrix}
= \begin{bmatrix}
0 \\
\vdots \\
0
\end{bmatrix}, \label{matrix translation, 2v-1, second case}
\end{align}
The linear system of equations in \eqref{matrix translation, 2v-1, second case} has
\begin{align*}
    (m-1)v+(\beta-m+1)(2v-1)+(m-1)v+(h-m+1)=\beta(2v-1)+h=t
\end{align*}
variables and $t=\beta(2v-1)+h$ equations. Recall that we want to show $\Delta_{\beta+m}=\dots=\Delta_K=0$. \\
\textbf{Step $\mathbf{3}$}. Let 
\begin{align*}
\mathbf{B}\triangleq[\mathbf{B}_1\mathbf{E}_1|\dots|\mathbf{B}_{m-1}\mathbf{E}_{m-1}|\mathbf{B}_{m}\mathbf{D}_{m} | \dots|\mathbf{B}_{\beta}\mathbf{D}_{\beta}|\mathbf{B}_{\beta+1}\mathbf{E}_{\beta+1}|\dots|\mathbf{B}_{\beta+m-1}\mathbf{E}_{\beta+m-1}],
\end{align*}
and $\textrm{rank}(\mathbf{B})=r$ for some $r\in \mathbb{N}$. Since the size of $\mathbf{B}$ is $t\times (2\beta v-\beta+m-1)$, we have $r\le 2\beta v-\beta+m-1$ (recall that $m\le h$, so $2\beta v-\beta+m-1\le 2\beta v-\beta+h=t$, and $\mathbf{B}$ is a tall matrix). For a full rank submatrix of $\mathbf{B}$ of size $t\times r$, containing the $r$ linearly independent columns of $\mathbf{B}$, which we denote by $\mathbf{\tilde{B}}$, there exists a variable vector $[w'_1,\dots,w'_r]$, such that
\begin{align} \label{A', second case}
\sum_{k=1}^{m-1}\mathbf{B}_k\mathbf{E}_k \begin{bmatrix}
w_{1,k} \\
\vdots \\
w_{v,k}
\end{bmatrix} +
\sum_{k=m}^{\beta}\mathbf{B}_k\mathbf{D}_k
\begin{bmatrix}
w_{1,k} \\
\vdots\\
w_{2v-1,k}
\end{bmatrix} + \sum_{k=\beta+1}^{\beta+m-1}\mathbf{B}_k\mathbf{E}_k \begin{bmatrix}
w_{1,k} \\
\vdots \\
w_{v,k}
\end{bmatrix} = \mathbf{\tilde{B}} \begin{bmatrix}
w'_1 \\
\vdots \\
w'_r
\end{bmatrix}.
\end{align}
Using \eqref{A', second case}, we rewrite \eqref{matrix translation, 2v-1, second case} as
\begin{align} \label{final linear, second case}
[\mathbf{\tilde{B}}|\mathbf{C}]\begin{bmatrix}
w'_1 \\
\vdots \\
w'_r \\
\Delta_{\beta+m} \\
\vdots \\
\Delta_K
\end{bmatrix} = \begin{bmatrix}
0\\
\vdots\\
0
\end{bmatrix}.
\end{align}
\textbf{Step $\mathbf{4}$}. The size of $[\mathbf{\tilde{B}}|\mathbf{C}]$ is $t\times(r+h-m+1)$. Because $r\le 2\beta v-\beta+m-1$, or $r+h-m+1\le h+2\beta v-\beta =t$, $[\mathbf{\tilde{B}}|\mathbf{C}]$ is a tall matrix. This matrix is full rank with high probability, and the proof is similar to the proof in Step $3$ in the previous case. Therefore, the unique solution of \eqref{final linear, second case} is $w'_1=\dots=w'_r=\Delta_{\beta+m}=\dots=\Delta_K=0$, which means $z_k=x_k$ for $k\in\mathcal{\hat{H}}\cap\mathcal{H}$.

\large \textbf{Case$\mathbf{3}$. $\mathbf{\mathcal{\hat{A}}\cap\mathcal{A}=\varnothing}$.}\\ \normalsize
 \textbf{Step $\mathbf{1}$}. Without loss of generality, we assume that $\mathcal{\hat{A}}=\{\beta+1,\dots,2\beta\}$, and $2\beta\le K$. We have the following linear system of equations.
\begin{align*}
    L_n(z_1,\dots,z_{\beta},z_{n,\beta+1},\dots,z_{n,2\beta},x_{2\beta+1}+\Delta_{2\beta+1},\dots,x_K+\Delta_K) =  L_n(x_{n1},\dots,x_{n\beta},x_{\beta+1},\dots,x_K), n\in \mathcal{T},
\end{align*}
where $\{z_{n,k}\}_{n\in \mathcal{T}}=\{z_k^{(i)}\}_{i\in[v]}$ for $k\in \mathcal{\hat{A}}$, and $\{x_{nk}\}_{n\in \mathcal{T}}=\{x_k^{(i)}\}_{i\in[v]}$ for $k\in[\beta]$. In addition, those equations are linearly independent and the adversary cannot make them dependent by exploiting its options.\\
The above system of equations is equivalent to
\begin{align}
L_n(z_1-x_{n,1},\dots,z_{\beta}-x_{n,\beta},z_{n,\beta+1}-x_{\beta+1},\dots,z_{n,2\beta}-x_{2\beta},\Delta_{2\beta+1},\dots,\Delta_K) =  0, \quad n\in \mathcal{T}. \label{case3 the new linear system}
\end{align}
This system of equations can be expressed as
\begin{align}
\sum_{k=1}^{\beta}\mathbf{B}_k \begin{bmatrix}
z_k-x_{n_1,k} \\
\vdots \\
z_k-x_{n_t,k}
\end{bmatrix} + \sum_{k=\beta+1}^{2\beta}\mathbf{B}_k \begin{bmatrix}
z_{n_1,k}-x_k \\
\vdots \\
z_{n_t,k}-x_k
\end{bmatrix} + \mathbf{C}
\begin{bmatrix}
\Delta_{2\beta+1} \\
\vdots \\
\Delta_K
\end{bmatrix}
= \begin{bmatrix}
0 \\
\vdots \\
0
\end{bmatrix}, \label{case3 matrix translation}
\end{align}
where $\mathbf{B}_k$, $k\in [2\beta]$, and $\mathbf{C}$ are $t\times t$ and $t\times (K-2\beta)$ matrices respectively, and are composed of the coefficients of the linear code.\\ \textbf{Step $\mathbf{2}$}. Since \[\{z_k-x_{nk}\}_{n\in \mathcal{T}}=\{z_k-x_k^{(i)}\}_{i\in[v]}\triangleq \{w_{i,k}\}_{i\in[v]},\quad k\in[\beta],\] and 
\begin{align*}
\{z_{n,k}-x_k\}_{n\in \mathcal{T}} =  \{z_k^{(i)}-x_k\}_{i\in[v]} \triangleq \{w_{i,k}\}_{i\in[v]}, \quad \beta+1\le k\le 2\beta,
\end{align*}
there exist $t\times v$ constant matrices $\mathbf{D}_k$ such that 
\begin{align}
\begin{bmatrix}
z_k-x_{n_1,k} \\
\vdots\\
z_k-x_{n_t,k}
\end{bmatrix} = \mathbf{D}_k \begin{bmatrix}
w_{1,k} \\
\vdots \\
w_{v,k}
\end{bmatrix}, \quad k\in[\beta],
\end{align}
and 
\begin{align}
\begin{bmatrix}
z_{n_1,k}-x_k \\
\vdots\\
z_{n_t,k}-x_k
\end{bmatrix} = \mathbf{D}_k \begin{bmatrix}
w_{1,k} \\
\vdots \\
w_{v,k}
\end{bmatrix}, \quad \beta+1\le k\le 2\beta.
\end{align}
Therefore, \eqref{case3 matrix translation} can be rewritten as
\begin{align}
\sum_{k=1}^{\beta}\mathbf{B}_k\mathbf{D}_k \begin{bmatrix}
w_{1,k} \\
\vdots \\
w_{v,k}
\end{bmatrix} + \sum_{k=\beta+1}^{2\beta}\mathbf{B}_k\mathbf{D}_k \begin{bmatrix}
w_{1,k} \\
\vdots \\
w_{v,k}
\end{bmatrix} + \mathbf{C}
\begin{bmatrix}
\Delta_{2\beta+1} \\
\vdots \\
\Delta_K
\end{bmatrix}
= \begin{bmatrix}
0 \\
\vdots \\
0
\end{bmatrix}, \label{case3 matrix translation, 2v-1}
\end{align}
The linear system of equations in \eqref{case3 matrix translation, 2v-1} has
\begin{align*}
    \beta v+\beta v+(h-\beta)=\beta(2v-1)+h
\end{align*}
variables and $t=\beta(2v-1)+h$ equations. \\
\textbf{Step $\mathbf{3}$}. Let 
\begin{align*}
\mathbf{B}\triangleq[\mathbf{B}_1\mathbf{D}_1|\dots|\mathbf{B}_{\beta}\mathbf{D}_{\beta}|\mathbf{B}_{\beta+1}\mathbf{D}_{\beta+1}|\dots|\mathbf{B}_{2\beta}\mathbf{D}_{2\beta}],
\end{align*}
and $\textrm{rank}(\mathbf{B})=r$ for some $r\in \mathbb{N}$. Since size of $\mathbf{B}$ is $t\times (2\beta v)$, we have $r\le 2\beta v$. For a full rank submatrix of $\mathbf{B}$ of size $t\times r$, containing the $r$ linearly independent columns of $\mathbf{B}$, which we denote by $\mathbf{\tilde{B}}$, there exists a variable vector $[w'_1,\dots,w'_r]$, such that
\begin{align} \label{case3 A'}
\sum_{k=1}^{\beta}\mathbf{B}_k\mathbf{D}_k \begin{bmatrix}
w_{1,k} \\
\vdots \\
w_{v,k}
\end{bmatrix} + \sum_{k=\beta+1}^{2\beta}\mathbf{B}_k\mathbf{D}_k \begin{bmatrix}
w_{1,k} \\
\vdots \\
w_{v,k}
\end{bmatrix} = \mathbf{\tilde{B}} \begin{bmatrix}
w'_1 \\
\vdots \\
w'_r
\end{bmatrix}.
\end{align}
Using \eqref{case3 A'}, we rewrite \eqref{case3 matrix translation, 2v-1} as
\begin{align} \label{case3 final linear}
[\mathbf{\tilde{B}}|\mathbf{C}]\begin{bmatrix}
w'_1 \\
\vdots \\
w'_r \\
\Delta_{2\beta+1} \\
\vdots \\
\Delta_K
\end{bmatrix} = \begin{bmatrix}
0\\
\vdots\\
0
\end{bmatrix}.
\end{align}
\textbf{Step $\mathbf{4}$}. The size of $[\mathbf{\tilde{B}}|\mathbf{C}]$ is $t\times(r+h-\beta)$. Because $r\le 2\beta v$, or $r+h-\beta\le h+2\beta v-\beta =t$, $[\mathbf{\tilde{B}}|\mathbf{C}]$ is a tall matrix. This matrix is full rank with high probability, and the proof is similar to the proof in Step $3$ in the first case. Therefore, the unique solution of \eqref{case3 final linear} is $w'_1=\dots=w'_r=\Delta_{2\beta+1}=\dots=\Delta_K=0$.
\end{proof}

\section{Converse proof for $t^*_{\textrm{linear}}(N,K,\beta,v)$} \label{linear converse section}
In this section, we show that for any $t$-correct linear code for a $(N,K,\beta,v)$ distributed system with linear functions $f_1,\dots,f_N$, if $t<t^*_{\textrm{Linear}}$,  there exist a set $\mathcal{T}\subseteq[N]$, $|\mathcal{T}|=t$, and a valid assignment of $\{x_{nk}\}_{n\in\mathcal{T},k\in[K]}$, that results in two feasible solutions that differ in the message of at least one honest node. It suffices to show this for $t=t^*_{\textrm{Linear}}-1$. The idea in the proof is that the adversary can act in a particular way so that in \textit{any} linear code, the linear system of $t$ equations in the decoder has $h+2\beta v-\beta$ variables, and are linearly independent. Since $t<h+2\beta v-\beta$, i.e. the number of the variables is more than the number of equations, this causes ambiguity about the messages of the honest nodes. In other words, for at least one $k\in\mathcal{H}$, we would have $\hat{x}_k\neq x_k$. Note that, naturally, there is no unique solution to an underdetermined system, i.e. one with more variables than equations. But here, we need more than only a condition on the number of equations and variables. This is because we want to show that in such underdetermined system, there is ambiguity about some certain variables, more specifically, those for the honest nodes. For example, in some codes, all the other variables might fall in a separate subspace, so that the honest variables can be uniquely found. In the converse proof, we show that no such thing occurs, and in any linear $t$-correct code, $t^*_{\textrm{Linear}}-1$ equations are always insufficient for unique decoding of the messages of the honest nodes.

In the following, first we state a definition and two lemmas and then explain how to find such assignment, in the two regions $K\le N\le K+2\beta(v-1)$ and $N\ge K+2\beta(v-1)$. These lemmas are for the general distributed encoding system and are not limited to the linear regime.

\begin{definition}
A multivariable function $f(x_1,\dots,x_S): \mathbb{F}^S\rightarrow \mathbb{F}$, $S\in\mathbb{N}$, is said to be a function of the variable $x_1$, if there exist $x_2,\dots,x_S,\delta\in\mathbb{F}$, such that $f(x_1+\delta,x_2\dots,x_S)\neq f(x_1,x_2,\dots,x_S)$. Functionality of the other variables $x_2,\dots,x_K$ is defined similarly.
\end{definition}
If $f(x_1,\dots,x_S)$ is a linear function, i.e. $f(x_1,\dots,x_S)=\sum_{i\in[S]} \alpha_i x_i$, the above definition states that $f$ is a function of $x_1$, if $\alpha_1\neq 0$.

\begin{lemma}\label{zero lemma}
Consider a $t$-correct code for a general $(N,K,\beta,v)$ distributed encoding system with the encoding functions $f_1(x_1,\dots,x_K),\dots,f_N(x_1,\dots,x_K)$. For any $m$ arbitrary subset of the encoding functions, namely $f_{n_1},\dots,f_{n_m}$, $m\in [K]$, there are at least $m$ input symbols $x_{k_1}, \ldots, x_{k_m}$  out of the $K$ input symbols, such that for each symbol $x_{k_j}$, $j \in [m]$, at least one  of $f_{n_1},\dots,f_{n_m}$ is a function of $x_{k_j}$. 
\end{lemma}
This lemma is only a consequence of the MDS property of a $t$-correct code. 
\begin{proof}
 The case $m=1$ is trivial, because a function $f_{n_1}$ is a function of one or more variables. The case $m=K$ immediately follows from the MDS property of $t$-correct codes. For $2\le m\le K-1$, by contradiction suppose that there are at most $m-1$ inputs, such that each of $f_1,\dots,f_m$ is only a function of some or all of them. Without loss of generality, let those $m-1$ inputs be $x_1,\dots,x_{m-1}$. According to the MDS property, the values of $f_1,\dots,f_K$ determine $x_1,\dots, x_K$ uniquely. None of $f_1,\dots,f_m$ is a function of any of $x_m,\dots,x_K$, so the values of $f_{m+1},\dots,f_K$ should determine $x_m,\dots,x_K$ uniquely, . But this is impossible, because there is no bijective mapping from $\mathbb{F}^{K-m+1}$ to $\mathbb{F}^{K-m}$.
\end{proof}

\begin{lemma} \label{first lemma}
Consider a $t$-correct code for a general $(N,K,\beta,v)$ distributed encoding system with the encoding functions $f_1(x_1,\dots,x_K),\dots,f_N(x_1,\dots,x_K)$. If for \emph{any} $\beta$ source nodes $\{k_1\dots,k_{\beta}\} \subset [K]$, there are $h=K-\beta$ functions $f_{n_1},\dots,f_{n_{h}}$,  that are \emph{not} functions of any of $x_{k_1}\dots,x_{k_{\beta}}$, then there are exactly $K$ univariate functions $f_{i_1},\dots,f_{i_K}$, such that $f_{i_k}$
is only a function of $x_{i_k}$, for $k\in[K]$.
\end{lemma}
Consider an example in which $f_i(x_1,\dots,x_K)=g_i(x_i)$, $g_i:\mathbb{F}\rightarrow\mathbb{F}$, for $i\in[K]$, i.e. the first $K$ functions $f_1,\dots,f_K$ are univariate. Now, for a subset of $[K]$ of size $\beta$, say $\mathcal{S}=\{k_1,\dots,k_{\beta}\}$, the $h$ functions $f_k, k\in[K]\setminus\mathcal{S}$, are not functions of any of $x_k, k\in\mathcal{S}$. Lemma \ref{first lemma} states that the reverse statement holds as well: If for any $\beta$ inputs, there are $h$ functions with the said property, there would be $K$ univariate functions among $f_1,\dots,f_N$. The emphasis that there are exactly $K$ such functions is due to the MDS property, because two functions of only one variable violates Lemma \ref{zero lemma} for $m=2$.
\begin{proof}
The proof of this lemma can be found in Appendix \ref{appendix 1}.
\end{proof}
Now we get back to the converse proof. First, we provide the proof for the region $N\ge h+2\beta v-\beta$. 
 \subsection{Converse proof for $N\ge h+2\beta v-\beta$}
\textbf{Step $\mathbf{1}$}. This step is different for $\beta=1$ and $\beta\ge 2$. First let $\beta=1$. Since $N\geq h+2v-1$, there are at least $2v-1$ functions in $f_1,\dots,f_N$, that are functions of $x_1$, according to Lemma \ref{zero lemma}. We choose $h+2v-2$ functions that include at least $2v-1$ functions of $x_1$. Without loss of generality, suppose they are $f_1,\dots,f_{h+2v-2}$. Therefore, there are at most $h-1$ functions among $f_1,\dots,f_{h+2v-2}$ that are not functions of $x_1$.\\
Now let $\beta\ge2$. Due to the MDS property of any $t$-correct code, there is at most one univariate function of each input among $f_1,\dots,f_N$. Since $N\geq h+2\beta v-\beta$, we choose $t=h+2\beta v-\beta-1$ functions out of $f_1,\dots,f_N$ that include at most $K-1$ univariate functions. Without loss of generality, suppose that they are $f_1,\dots,f_t$. According to Lemma \ref{first lemma}, there exist $\beta$ source nodes such that at most $h-1$ function out of $f_1,\dots,f_t$ are not functions of them. Without loss of generality, suppose that they are the first $\beta$ source nodes. The following proposition summarizes this step.
\begin{proposition} \label{proposition 1}
We can choose $t=h+2\beta v-\beta -1$ functions out of $f_1,\dots,f_N$, without loss of generality $f_1,\dots,f_t$, such that at most $h-1$ functions out of them are not functions of the first $\beta$ inputs.
\end{proposition}
We choose the first $\beta$ source nodes as the adversaries, and use the above proposition in the last step of the proof.\\
\textbf{Step $\mathbf{2}$}. Our goal is to find $\{x_k^{(i)}\}_{i\in[v], k\in[\beta]}$, $\{z_k^{(i)}\}_{i\in[v], k\in[\beta]}$, $\{x_k\}_{\beta+1\le k\le K}$, and $\{\Delta_k\}_{\beta+1\le k\le K}$, such that
\begin{align}
    f_n(x_{n1},\dots,x_{n\beta},x_{\beta+1},\dots,x_K) = f_n(z_{n1},\dots,z_{n\beta},x_{\beta+1}+\Delta_{\beta+1},\dots,x_K+\Delta_K), \quad n\in[t], \label{f(x) = f(z)}
\end{align}
where $\{x_{nk}\}_{n\in[t]}=\{x^{(i)}_k\}_{i\in[v]}$, $\{z_{nk}\}_{n\in[t]}=\{z^{(i)}_k\}_{i\in[v]}$, for all $k\in[\beta]$, and at least one of $\Delta_k$, $\beta+1\le k\le K$ is nonzero. In other words, we want to find two setups of messages of the source nodes that make up the same codeword. The messages in setup $1$ are $x_{nk}$, $k\in[\beta]$, and $x_k$, $\beta+1\le k\le K$. The messages in setup $2$ are $z_{nk}$, $k\in[\beta]$, and $x_k+\Delta_k$, $\beta+1\le k\le K$, for $n\in[t]$. Equation \eqref{f(x) = f(z)} is equivalent to
\begin{align} \label{linear cpnverse objective}
f_n(z_{n1}-x_{n1},\dots,z_{n\beta}-x_{n\beta},\Delta_{\beta+1},\dots,\Delta_K)=0, \quad n\in[t].
\end{align}
We want the number of the variables in the above system to be $h+2\beta v-\beta$.

\textbf{Step $\mathbf{3}$}. We choose the configuration in table \eqref{linear converse assignment} for $k\in [\beta]$. This table signifies how adversaries send their messages to the encoding nodes in the two setups. For example, in setup $1$, the adversarial node $k\in[\beta]$ sends $x_1^{(1)}$ to encoding nodes $n\in \mathcal{N}_1$, and in setup $2$ sends $z_1^{(1)}$ to encoding nodes $n\in \mathcal{N}_1\cup \mathcal{N}_2$. 

\begin{eqnarray}{
\begin{array}{|c|c|c|}
\hline
 n                          &  x_{nk}                      &  z_{nk}                      \\ \hline
 \mathcal{N}_1=\{1,\dots,h+\beta-1\}                    &  x_k^{(1)}                   & \multirow{2}{*}{$z_k^{(1)}$} \\ \cline{1-2}
 \mathcal{N}_2=\{h+\beta,\dots,h+2\beta-1\}             & \multirow{2}{*}{$x_k^{(2)}$} &                              \\ \cline{1-1} \cline{3-3} 
 \mathcal{N}_3=\{h+2\beta,\dots,h+3\beta-1\}            &                              & \multirow{2}{*}{$z_k^{(2)}$} \\ \cline{1-2}
 \mathcal{N}_4=\{h+3\beta,\dots,h+4\beta-1\}            &  x_k^{(3)}                   &                              \\ \hline
\multicolumn{3}{|c|}{\vdots}                                                                      \\ \hline
 \mathcal{N}_{2v-2}=\{h+(2v-3)\beta,\dots,h+(2v-2)\beta-1\}  & \multirow{2}{*}{$x_k^{(v)}$} &  z_k^{(v-1)}                 \\ \cline{1-1} \cline{3-3} 
 \mathcal{N}_{2v-1}=\{h+(2v-2)\beta,\dots,h+(2v-1)\beta-1\}  &                              &  z_k^{(v)}                   \\ \hline
\end{array} }
\label{linear converse assignment}
\end{eqnarray}

Let $w_{i+j-1,k}\triangleq z_k^{(i)}-x_k^{(j)}$, $i,j\in [v], k\in[\beta]$. We can rewrite \eqref{linear cpnverse objective} using \eqref{linear converse assignment} as following.
\begin{align}
\begin{split} 
f_n(w_{1,1},\dots,w_{1,\beta},\Delta_{\beta+1},\dots,\Delta_K)&=0, \quad n\in \mathcal{N}_1,  \\
f_n(w_{2,1},\dots,w_{2,\beta},\Delta_{\beta+1},\dots,\Delta_K)&=0, \quad n\in \mathcal{N}_2,  \\
f_n(w_{3,1},\dots,w_{3,\beta},\Delta_{\beta+1},\dots,\Delta_K)&=0, \quad n\in \mathcal{N}_3,  \\
f_n(w_{4,1},\dots,w_{4,\beta},\Delta_{\beta+1},\dots,\Delta_K)&=0, \quad n\in \mathcal{N}_4,  \\
 \quad \quad \vdots \\
f_n(w_{2v-2,1},\dots,w_{2v-2,\beta},\Delta_{\beta+1},\dots,\Delta_K)&=0, \quad n\in \mathcal{N}_{2v-2},  \\
f_n(w_{2v-1,1},\dots,w_{2v-1,\beta},\Delta_{\beta+1},\dots,\Delta_K)&=0, \quad n\in \mathcal{N}_{2v-1}.  \label{linear converse objective updated}
\end{split}
\end{align}
There are $h+2\beta v-\beta$ variables, $\{w_{i,k}\}_{i\in [2v-1],k\in[\beta]}$ and $\{\Delta_k\}_{\beta+1\le k\le K}$, in the above $t$ equations. We want these equations to be linearly independent. \\
\textbf{Step $\mathbf{4}$}. Let \[\mathbf{w}\triangleq[\mathbf{w}_1^T,\dots,\mathbf{w}_{2v-1}^T]^T,\]where $\mathbf{w}_i=[w_{i,1},\dots,w_{i,\beta}]^T$ for $i\in [2v-1]$, and $\mathbf{\Delta}=[\Delta_{\beta+1},\dots,\Delta_K]^T$.
Assume that the generator matrix of the linear code is $\mathbf{G}=[g_{n,k}]_{n\in [N],k\in [K]}$, i.e. 
\begin{align}
    f_n(x_{n1},\dots,x_{nK})=\sum_{k=1}^{K}g_{n,k}x_{nk}, \quad n \in [N].
\end{align}
Therefore, the matrix form of \eqref{linear converse objective updated} is
\begin{eqnarray}{ \mathbf{B} \begin{bmatrix}
\mathbf{w} \\
\mathbf{\Delta}
\end{bmatrix} =\left[
\begin{array}{ccc|c}
\mathbf{C}_1     &       &      & \multirow{4}{*}{$\mathbf{D}$} \\
      & \mathbf{C}_2     &      &                    \\
\multicolumn{3}{c|}{\ddots} &                    \\
      &       & \mathbf{C}_{2v-1}    &    
\end{array} \right] \begin{bmatrix}
\mathbf{w} \\
\mathbf{\Delta}
\end{bmatrix} =\begin{bmatrix}
0\\ 
\vdots \\
0
\end{bmatrix},} \label{matrix form}
\end{eqnarray}
where 
\begin{align*}
    \mathbf{C}_i &= [g_{n,k}]_{n\in \mathcal{N}_i,k\in[\beta]}, \quad i\in [2v-1], \\
    \mathbf{D} &= [g_{n,k}]_{n\in [t], \beta+1\le k\le K},
\end{align*}
and the unspecified elements in \eqref{matrix form} are zeros. The sizes of the submatrices in \eqref{matrix form} are as follows.
\begin{eqnarray}{
\begin{array}{|c|c|}
\hline
\textrm{Matrix} & \textrm{Size} \\ \hline
 \mathbf{C}_1 &  (h+\beta-1) \times \beta \\ \hline
\mathbf{C}_2,\dots,\mathbf{C}_{2v-1}  &  \beta\times\beta\\ \hline
 \mathbf{D} &  (h+2\beta v-\beta-1) \times h\\ \hline
\mathbf{B} &  (h+2\beta v-\beta-1)\times(h+2\beta v-\beta) \\ \hline
\end{array}}
\end{eqnarray}
Therefore, $\textrm{rank}(\mathbf{B})\le h+2\beta v-\beta-1$. \\
\textbf{Step $\mathbf{5}$}. The generator matrix $\mathbf{G}$ is full, and because $\mathbf{D}$ is composed of $h$ columns of it (in the first $t$ rows), $\mathbf{D}$ is full rank as well. The submatrix 
\begin{eqnarray}{ \mathbf{B^*} =\left[
\begin{array}{ccc}
\mathbf{C}_1     &       &      \\
      & \mathbf{C}_2     &                    \\
\multicolumn{3}{c}{\ddots}         \\
      &       & \mathbf{C}_{2v-1}      
\end{array} \right], } \nonumber
\end{eqnarray}
of $\mathbf{B}$ is full rank, if and only if all $\{\mathbf{C}_i\}_{i\in [2v-1]}$ are full rank. In Lemma \ref{full rank A*} that comes later, we show that if $\mathbf{B^*}$ is not full rank, it can be made full rank by modifying $\mathcal{N}_1,\dots,\mathcal{N}_{2v-1}$ in \eqref{linear converse assignment}. Thus, we consider $\mathbf{B^*}$ to be full rank. This means the $t$ equation in \eqref{linear converse objective updated} are linearly independent.

Since $\textrm{rank}(\mathbf{B})\le h+2\beta v-\beta-1$, and $\textrm{size}(\mathbf{B})=(h+2\beta v-\beta-1)\times(h+2\beta v-\beta)$, the columns of $\mathbf{B}$ are linearly dependent. Thus, there is a linear combination of the columns of $\mathbf{B}$ that equals zero. But, $\mathbf{B^*}$ and $\mathbf{D}$ are both full column rank, so this combination contains at least one column from $\mathbf{B^*}$ and one column from $\mathbf{D}$. Therefore, \eqref{matrix form} has a nonzero solution in which $\mathbf{\Delta}\neq \mathbf{0}$. It shows that the two setups in \eqref{f(x) = f(z)} correspond to the same codeword, even though the message of at least one honest node is different in them. This completes the converse proof. In the following, we state the lemma that allowed us to assume $\mathbf{B}^*$ is full rank.

\begin{lemma} \label{full rank A*}
Consider the $(h+2\beta v-\beta-1)\times\beta$ matrix $\mathbf{E}=[e_{n,k}]_{n\in [h+2\beta v-\beta-1],k\in[\beta]}$ with following properties.
\begin{itemize}
    \item $\mathbf{E}$ is full rank.
    \item $\mathbf{E}$ has $h-1$ zero rows at most.
    \item The submatrix consisting of any $h+\beta$ rows of $\mathbf{E}$ is full rank.
\end{itemize}
There is a partitioning $\{1,\dots, h+2\beta v-\beta-1\}=\mathcal{R}_1\cup\dots\cup\mathcal{R}_{2v-1}$, such that $|\mathcal{R}_1|=h+\beta-1$, $|\mathcal{R}_i|=\beta$ for $2\le i\le 2v-1$, and the submatrices $\mathbf{B}_j=[e_{n,k}]_{n\in \mathcal{R}_i,k\in [\beta]}$, $j\in [2v-1]$ are full rank.  
\end{lemma}
This lemma states that we can divide $\mathbf{E}$ row-wise, into $2v-1$ full rank submatrices. In order to use it to show that $\mathbf{B}^*$ is full rank, we need to verify that $\mathbf{E}=[g_{n,k}]_{n\in [t],k\in[\beta]}$ has the properties stated in the lemma. Recall that $\mathbf{G}=[g_{n,k}]_{n\in[N],k\in[K]}$ is the generator matrix of the linear code.
\begin{itemize}
    \item 
    Property 1: We know that the submatrix consisting of any $t$ rows from $\mathbf{G}$ is full rank. Therefore, $\mathbf{E}$ is full rank.
    \item
    Property 2: According to Proposition \ref{proposition 1}, at most $h-1$ functions in $f_1,\dots,f_t$ are not functions of the first $\beta$ inputs. Therefore, at most $h-1$ rows of $\mathbf{E}$ are zeros. 
    \item
    Property 3: According to the MDS property of $t$-codes, the submatrix of any $K=h+\beta$ rows of $\mathbf{G}$ is full rank, thus the submatrix of any $h+\beta$ rows of $\mathbf{E}$ is also full rank.
\end{itemize}
Now we present the proof of the lemma
\begin{proof}
 The case $v=1$ is trivial. So suppose $v\ge 2$. Let $r_n$ denote the $n$'th row of $\mathbf{E}$, where $n\in [t]$. We know that according to property 3, there are $\beta$ linearly independent rows among the first $h+\beta$ rows of $\mathbf{E}$. Let $\mathcal{R}'_2$ be the set containing the indices of those rows. There are $\beta$ independent rows among the first $h+\beta$ rows of $\mathbf{E}-\{r_n\}_{n\in \mathcal{R}'_2}$. Let $\mathcal{R}'_3$ be the set containing the indices of those rows. Similarly, there are $\beta$ independent rows among the first $h+\beta$ rows of $\mathbf{E}-\{r_n\}_{n\in \mathcal{R}'_2}-\{r_n\}_{n\in \mathcal{R}'_3}$. In the same way, we construct $\mathcal{R}'_2,\dots,\mathcal{R}'_{2v-1}$. In the last step, $h+\beta-1$ unchosen rows from $\mathbf{C}$ remain, by which we construct $\mathcal{R}'_1$. If $\{r_n\}_{n\in \mathcal{R}'_1}$ are full rank, the proof is complete. In the case of $\beta=1$, the $h$ rows in $\{r_n\}_{n\in \mathcal{R}'_1}$ are full rank, because each row is one element, and according to property 3, there are at most $h-1$ zero rows. Consequently, for $\beta=1$, the partitioning $\{1,\dots, h+2\beta v-\beta-1\}=\mathcal{R}_1\cup\dots\cup\mathcal{R}_{2v-1}$, where $\mathcal{R}_i=\mathcal{R}'_i$, $i\in [2v-1]$, satisfies the lemma.

Suppose that $\beta\ge 2$, and the row set $\{r_n\}_{n\in \mathcal{R}'_1}$ is not full rank. According to property 3, adding any other row to $\{r_n\}_{n\in \mathcal{R}'_1}$ makes it full rank. Thus, the rank of $\{r_n\}_{n\in \mathcal{R}'_1}$ is $\beta-1$, and there are $\beta-1$ independent rows in $\{r_n\}_{n\in \mathcal{R}'_1}$. 

We claim that we can construct $\mathcal{R}_1$ and $\mathcal{R}_2$ by replacing a row from $\{r_n\}_{n\in \mathcal{R}'_1}$, with a row from $\{r_n\}_{n\in \mathcal{R}'_2}$, such that $\{r_n\}_{n\in \mathcal{R}_1}$ and $\{r_n\}_{n\in \mathcal{R}_2}$ are full rank. We choose a nonzero row from $\{r_n\}_{n\in \mathcal{R}'_1}$, other than the $\beta-1$ aforementioned independent rows, and denote it by $r^*$. Since there are at most $h-1$ zero rows in $\{r_n\}_{n\in \mathcal{R}'_1}$, such row exists. Let $\mathcal{R}'_2=\{n_1,\dots,n_{\beta}\}\subset \mathbb{N}^{\beta}$. There is a row in $\{r_n\}_{n\in \mathcal{R}'_2}$, which we denote by $\hat{r}$, such that $r^*$ is not a linear combination of the rows $\{r_n\}_{n\in \mathcal{R}'_2}-\hat{r}$. In order to show this, by contradiction suppose that there exist coefficients $\sigma_{i,j}$, $i\in[\beta],j\in[\beta]\setminus\{i\}$, such that
\begin{align}
    r^* = \sum_{\substack{k\in [\beta] \\ k \neq 1}} \sigma_{1,k} r_{n_k} = \sum_{\substack{k\in [\beta] \\ k \neq 2}} \sigma_{2,k} r_{n_k} = \dots = \sum_{\substack{k\in [\beta] \\ k \neq \beta}} \sigma_{\beta,k} r_{n_k}.
\end{align}
The equation
\begin{align*}
    \sum_{\substack{k\in [\beta] \\ k \neq 1}} \sigma_{1,k} r_{n_k} = \sum_{\substack{k\in [\beta] \\ k \neq i}} \sigma_{i,k} r_{n_k}, \quad 2\le i\le \beta,
\end{align*}
is equivalent to
\begin{align*}
    \sigma_{1,i}r_{n_i}-\sigma_{i,1}r_{n_1}+\sum_{\substack{k\in [\beta] \\ k \neq 1,i}} (\sigma_{1,k}-\sigma_{i,k}) r_{n_k}=0, \quad 2\le i\le \beta.
\end{align*}
Since $\{r_n\}_{n\in \mathcal{R}'_2}$ is full rank, this means $\sigma_{1,k}=0$ for $2\le k\le\beta$. Consequently $r^*=0$, which is wrong. Therefore, $\hat{r}\in \{r_n\}_{n\in \mathcal{R}'_2}$ exists such that $r^*$ is not a linear combination of $\{r_n\}_{n\in \mathcal{R}'_2}-\hat{r}$. 

 Now we exchange $r^*$ from $\mathcal{R}'_1$ with $\hat{r}$ from $\mathcal{R}'_2$. Since $\{r_n\}_{n\in \mathcal{R}'_1}\cup\{\hat{r}\}$ is full rank, according to property 3, $\{r_n\}_{n\in \mathcal{R}'_1}\cup\{\hat{r}\}-\{r^*\}$ is full rank as well. Therefore, with this exchange, $\{r_n\}_{n\in \mathcal{R}_1}=\{r_n\}_{n\in \mathcal{R}'_1}\cup\{\hat{r}\}-\{r^*\}$ and $\{r_n\}_{n\in \mathcal{R}_2}=\{r_n\}_{n\in \mathcal{R}'_2}\cup\{r^*\}-\{\hat{r}\}$ are both full rank. So we found a partitioning $\{1,\dots, h+2\beta v-\beta-1\}=\mathcal{R}_1\cup\dots\cup\mathcal{R}_{2v-1}$, where $\mathcal{R}_i=\mathcal{R}'_i$ for $3\le i\le 2v-1$, that satisfies the lemma.
\end{proof}
\subsection{Converse proof for $K\le N< h+2\beta v-\beta$}
We want to prove $t^*_{\textrm{Linear}}=N$, so we show that $t=N-1$ results in two feasible solutions that differ in the message of at least one honest node. First let $\beta=1$. There are at least $N-h$ functions in $f_1,\dots,f_N$, that are functions of $x_1$, according to Lemma \ref{zero lemma}. We choose $t=N-1$ functions that include at least $N-h$ functions of $x_1$. Without loss of generality, suppose they are $f_1,\dots,f_{N-1}$. Therefore, there are at most $h-1$ functions among $f_1,\dots,f_{N-1}$ that are not functions of $x_1$.\\
Now let $\beta\ge2$. Due to the MDS property of any $t$-correct code, there is at most one univariate function of each input among $f_1,\dots,f_N$. We choose $t=N-1$ functions out of $f_1,\dots,f_N$ that include at most $K-1$ univariate functions. Without loss of generality, suppose that they are $f_1,\dots,f_{N-1}$. According to Lemma \ref{first lemma}, there exist $\beta$ source nodes, where at most $h-1$ function out of $f_1,\dots,f_{N-1}$ are not functions of them. Without loss of generality, suppose that they are the first $\beta$ source nodes.\\
The summary of the results do far is that 
we can choose $t=N-1$ functions out of $f_1,\dots,f_N$, without loss of generality $f_1,\dots,f_{N-1}$, such that at most $h-1$ functions out of them are not functions of the first $\beta$ inputs.\\
We choose the first $\beta$ source nodes as adversaries. Our goal is to find $\{x_k^{(i)}\}_{i\in[v], k\in[\beta]}$, $\{z_k^{(i)}\}_{i\in[v], k\in[\beta]}$, $\{x_k\}_{\beta+1\le k\le K}$, and $\{\Delta_k\}_{\beta+1\le k\le K}$, such that
\begin{align}\label{region 2}
    f_n(x_{n1},\dots,x_{n\beta},x_{\beta+1},\dots,x_K) = f_n(z_{n1},\dots,z_{n\beta},x_{\beta+1}+\Delta_{\beta+1},\dots,x_K+\Delta_K), \quad n\in[N-1], 
\end{align}
where $\{x_{nk}\}_{n\in[N-1]}=\{x^{(i)}_k\}_{i\in[v]}$, $\{z_{nk}\}_{n\in[N-1]}=\{z^{(i)}_k\}_{i\in[v]}$, for all $k\in[\beta]$, and at least one of $\Delta_k$, $\beta+1\le k\le K$ is nonzero. In the previous subsection, we proved that $t<h+2\beta v-\beta$ equations are not enough for the decoder, because they always cause ambiguity in the message of at least one honest node. Here, we have $N-1<h+2\beta v-\beta$ equations in \eqref{region 2}. Therefore, the result of the previous subsection applies here, and concludes the proof for $K\le N< h+2\beta v-\beta$.

\section{Discussion and Concluding Remarks} \label{discussion section}
We introduced the problem of adversarial distributed encoding, wherein $K$ source nodes send their data symbols to $N$ encoding nodes, that each calculates and stores an encoded symbol of the received symbols. Meanwhile, an adversary controls $\beta$ source nodes, where each sends $v$ different symbols to the encoding nodes. We study the fundamental limit of this system, $t^*(N,K,\beta,v)$, which is the minimum number of the encoding nodes that are needed to recover the messages of the honest nodes correctly. In the linear regime, i.e. when encoding nodes use linear functions, we characterized the fundamental limit $t^*_{\textrm{Linear}}$. 

There are many more aspects of distributed encoding systems that can be explored. Some of the potential research areas are as follows.
\subsection{Characterizing $t^*$}
We conjecture that $t^*< t^*_{\textrm{linear}}$ for large enough $N$, and that nonlinear coding is required in order to achieve $t^*$. Designing a $t$-correct nonlinear code, and proving the uniqueness of the messages of the honest nodes seems challenging because nonlinear systems usually have numerous solutions.
\subsection{Designing a linear code with efficient decoding algorithm}
Algorithm \ref{decoding alg} searches all the possible cases, looking for feasible solutions. Therefore, the decoding complexity is exponential, and such algorithm is not suitable for implementation. Very sparse codes or codes with algebraic structures may allow us to develop a low-complexity decoding algorithm. The following lemma shows that Reed-Solomon code achieves ${\textrm{linear}}$.
\begin{lemma}\label{rs code lemma}
Reed-Solomon code achieves $t^*_{\textrm{linear}}$ in a $(N,K,\beta,v)$ distributed linear encoding system.
\end{lemma}
The proof can be found in Appendix \ref{appendix 2}.
\subsection{Finding the smallest finite field that produces a reasonable error probability.}
In this paper, we showed that if the field size is large, then with high probability, random linear code achieves $t^*_{\textrm{linear}}$. We can also use Shwartz-Zipple Lemma to show that if the field size is large enough, there exists a linear code that achieves $t^*_{\textrm{linear}}$. The question of interest is the minimum size of the field needed to achieve $t^*_{\textrm{linear}}$.

\appendix
\begin{appendices}
\section{Proof of Lemma \ref{first lemma}} \label{appendix 1}
In this appendix, we prove Lemma \ref{first lemma}.
For any two different sets $\mathcal{A}_1,\mathcal{A}_2\in [K]$, $|\mathcal{A}_1|=|\mathcal{A}_2|=\beta$, $\mathcal{A}_1 \cap \mathcal{A}_2 \neq \varnothing$, the $h$ functions that are not functions of $\{x_i\}_{i\in \mathcal{A}_1}$ and $\{x_i\}_{i\in \mathcal{A}_2}$ differ in at least one function. Otherwise, there are $h$ functions in $f_1,\dots,f_N$ that are functions of at most $h-1$ symbols out of the $K$ symbols (because $K-|\mathcal{A}_1\cup\mathcal{A}_2|\le h-1$), which contradicts Lemma \ref{zero lemma}.

Without loss of generality, assume for the first $\beta$ symbols, $x_1,\dots,x_{\beta}$, the $h$ functions $f_{\beta+1},\dots,f_{h+\beta}$ are not functions of them. Now take $x_2,\dots,x_{\beta+1}$. According to the above discussion, the $h$ functions that are not functions of them, differ with $f_{\beta+1},\dots,f_{h+\beta}$ in at least one function. Without loss of generality, suppose $f_1$ is that function. So $f_1$ is not a function of $x_2,\dots,x_{\beta+1}$, but is a function of $x_1$, because otherwise the $h+1$ functions $\{f_1,f_{\beta+1},\dots,f_{h+a}\}$ are not functions of $x_1,\dots,x_{\beta}$, which is against Lemma \ref{zero lemma}.
Now take $x_1x_3,\dots,x_{\beta+1}$. Again, the $h$ functions that are not functions of them, differ with $f_{\beta+1},\dots,f_{h+\beta}$ in at least one function. Without loss of generality, suppose $f_2$ is that function. So $f_2$ is not a function of $x_1x_3,\dots,x_{\beta+1}$, but is a function of $x_2$, because otherwise the $h+1$ functions $\{f_2,f_{\beta+1},\dots,f_{h+\beta}\}$ are not functions of $x_1,\dots,x_{\beta}$, which is against Lemma \ref{zero lemma}.

By similar arguments, we deduce that without loss of generality, $f_k$ is a function of $x_k$, and is not function of $\{x_i,x_{\beta+1}\}_{i\in[\beta]\setminus\{k\}}$, for $k\in[\beta]$. In the following, we show that $f_1,\dots,f_{\beta}$ are not functions of $x_{\beta+2},\dots,x_K$ as well. Take $x_2,\dots,x_{\beta},x_{\beta+2}$. We know that the $h$ functions that are not functions of them, differ with $f_{\beta+1},\dots,f_{h+\beta}$ in at least one function. This one different function is not among $f_2,\dots,f_{\beta}$, because they are functions of $x_2,\dots,x_{\beta}$, respectively. Suppose this function is among $f_{h+\beta+1},\dots,f_N$, and without loss of generality, suppose it is $f_{\beta+h+1}$. But the $h+2$ functions $\{f_1,f_{\beta+1},\dots,f_{\beta+h+1}\}$, are not functions of $x_2,\dots,x_{\beta}$, which is not possible due to Lemma \ref{zero lemma}. Therefore the only remaining choice is that $f_1$ is that different function that is not function of $x_2,\dots,x_{\beta},x_{\beta+2}$. 

So we proved that $f_1$ is a function of $x_1$, and is not a function of $x_2,\dots,x_{\beta+2}$. Similarly, $f_1$ is not a function of the remaining inputs, $x_{\beta+3},\dots,x_K$. Similarly, $f_2,\dots,f_{\beta}$ are not functions of $x_{\beta+2},\dots,x_K$. Therefor, $f_1,\dots,f_{\beta}$ are univariate functions of $x_1,\dots,x_{\beta}$, respectively. Since $x_1,\dots,x_{\beta}$ are arbitrary, we proved that there exists at least one univariate function of each of the $K$ input symbols. Due to the MDS property of $t$-codes, there is at most one univariate function of each input symbol. Therefore there are exactly $K$ univariate functions of $x_1,\dots,x_K$ among $f_1,\dots,f_N$.

\section{Proof of Lemma \ref{rs code lemma}} \label{appendix 2}
Recall that RS code is linear, and is described by the generator matrix $\mathbf{G}=[\alpha_{nk}]$ where $\alpha_{nk}=\lambda_n^{k-1}$, $n\in[N],k\in [\beta]$ and $\lambda_1,\dots,\lambda_N\in \mathbb{F}$ are distinct. In other words, 
\begin{align*}
    \mathbf{G} = \begin{bmatrix}
    1 & \lambda_1 & \lambda_1^2 & \dots & \lambda_1^{K-1} \\
    1 & \lambda_2 & \lambda_2^2 & \dots & \lambda_2^{K-1} \\
    & & \ddots & & \\
    1 & \lambda_N & \lambda_N^2 & \dots & \lambda_N^{K-1} 
    \end{bmatrix}.
\end{align*}
\begin{proof}[proof of Lemma \ref{rs code lemma}]
As the proof for the random linear code, suppose that the first $\beta$ source nodes are the adversaries, i.e. $\mathcal{A}=\{1,\dots, \beta\}$ and $\mathcal{H}=\{\beta+1,\dots,K\}$. The $v$ messages of the adversarial source node $k\in \mathcal{A}$ are $\{x_k^{(i)}\}_{i\in[v]}$, $k\in[\beta]$, and the message of the honest node $k\in \mathcal{H}$ is $x_k$, $\beta+1\le k\le K$. Assuming an arbitrary set $\mathcal{T}=\{n_1,\dots,n_t\}$, we want to show that the only partial solution for $f_n(x_{n1},\dots,x_{nK})=y_n$, $n\in \mathcal{T}$, is $x_{nk}=x_k$, $k\in\mathcal{H}$. 

Again, like the proof of achievability for the random linear code, we need to examine three cases $\mathcal{\hat{A}}=\{1,\dots,\beta\}$, $\mathcal{\hat{A}}=\{m,\dots,\beta+m-1\}$, $m\in[\beta]$, and $\mathcal{\hat{A}}=\{\beta+1,\dots,2\beta\}$. But since that proof showed that these three cases are similar, we only examine the first case $\mathcal{\hat{A}}=\{1,\dots,\beta\}$.

Suppose that in addition to the real scenario, there is another solution for $f_n(x_{n1},\dots,x_{nK})=y_n$, $n\in \mathcal{T}$ in which $\mathcal{\hat{A}}=\{1,\dots,\beta\}$, the messages of the source node $k\in \mathcal{\hat{A}}$ are $\{z_k^{(i)}\}_{i\in[v]}$, and the message of the source node $k\in \mathcal{\hat{H}}$ is $z_k=x_k+\Delta_k$. The goal is to prove $\Delta_k=0$ for $k\in \mathcal{\hat{H}}$.

Since RS code is linear, all the equations \eqref{first lienar} to \eqref{zeta linear combination} are the same here. Thus, we do not repeat them here and continue from \eqref{unique zeta}. We have
\begin{align}
    \begin{bmatrix}
    \zeta_1 \\
    \vdots \\
    \zeta_r
    \end{bmatrix} = \begin{bmatrix}
    \gamma_{n_1,1}\alpha_{n_1,1} & \dots & \gamma_{n_1,r}\alpha_{n_1,r} \\
    & \ddots & \\
    \gamma_{n_r,1}\alpha_{n_r,1} & \dots & \gamma_{n_r,r}\alpha_{n_r,r} 
    \end{bmatrix}^{-1} \begin{bmatrix}
    \alpha_{n_1,\beta+1} \\
    \vdots \\
    \alpha_{n_r,\beta+1}
    \end{bmatrix}= \begin{bmatrix}
    \gamma_{n_1,1}\lambda_{n_1}^0 & \dots & \gamma_{n_1,r}\lambda_{n_1}^{r-1} \\
    & \ddots & \\
    \gamma_{n_r,1}\lambda_{n_r}^0 & \dots & \gamma_{n_r,r}\lambda_{n_r}^{r-1} 
    \end{bmatrix}^{-1} \begin{bmatrix}
    \lambda_{n_1}^{\beta} \\
    \vdots \\
    \lambda_{n_r}^{\beta}
    \end{bmatrix}, \label{unique zeta rs}
\end{align}
and
\begin{align}
   \zeta_1\gamma_{n_{r+1},1}\lambda_{n_{r+1}}^0+\dots + \zeta_r\gamma_{n_{r+1},r}\lambda_{n_{r+1}}^{r-1} = \lambda_{n_{r+1}}^{\beta}. \label{linear zeta rs}
\end{align}
The above equation means that $\lambda_{n_{r+1}}$ is a root of the polynomial $g(x)=x^{\beta}-\zeta_r\gamma_{n_{r+1},r}x^{r-1}-\dots-\zeta_1\gamma_{n_{r+1},1}x^0$. But $\lambda_{n_{r+1}}$ is chosen uniformly at random from a large field, and is independent from $\lambda_{n_1},\dots,\lambda_{n_r}$ in \eqref{unique zeta rs}. Each option for $\{\gamma_{n_i,j}\}_{i\in[r+1],j\in[r]}\in\{0,1,-1\}^{r(r+1)}$ results in at most $\beta$ roots of $g(x)$ in the field, so the number of options that the adversary can try is limited. Given that $\lambda_1,\dots,\lambda_N$ are chosen uniformly at random from the field, the probability that the choice of the adversary satisfies \eqref{linear zeta rs} is negligible. After this, we can continue the proof for the linear code and prove that $[\mathbf{\tilde{B}}|\mathbf{C}]$ in \eqref{final linear} is full rank and conclude the proof.
\end{proof}

\end{appendices}

\bibliographystyle{IEEEtran}
\bibliography{ref}

\end{document}